\documentclass[12pt]{article}

\usepackage{amsmath}
\usepackage{amssymb}
\usepackage{amsthm}
\usepackage{hyperref}
\usepackage{enumerate}

\newtheorem{thm}{Theorem}[section]
\newtheorem{co}[thm]{Corollary}
\newtheorem{lem}[thm]{Lemma}

\newtheorem{assumption}[thm]{Assumption}

\newtheorem{definition}[thm]{Definition}

\newtheorem{example}[thm]{Example}

\newtheorem{remark}[thm]{Remark}
\newenvironment{rem}{\begin{remark}\rm}{\end{remark}}

\def\eps{\varepsilon}

\title{Limit Theorems in Hidden Markov Models}

\author{\begin{tabular}{cc}
Guangyue Han\\
University of Hong Kong\\
{\em email:} ghan@hku.hk\\
\end{tabular}}

\date{{\normalsize \today}}

\begin{document}\maketitle\thispagestyle{empty}

\begin{abstract}
In this paper, under mild assumptions, we derive a law of large numbers, a central limit theorem with an error estimate, an almost sure invariance principle and a variant of Chernoff bound in finite-state hidden Markov models. These limit theorems are of interest in certain ares in statistics and information theory. Particularly, we apply the limit theorems to derive the rate of convergence of the maximum likelihood estimator in finite-state hidden Markov models.
\end{abstract}

\section{Main Results and Related Work} \label{main-results}

Consider a discrete memoryless channel with a finite input alphabet $\mathcal{Y}$ and a finite output alphabet $\mathcal{Z}$. Assume that, at each time slot, the channel is characterized by the channel transition probability matrix $\Pi=(p(z|y))$. Let $Y=(Y_i: i \in \mathbb{Z})$ be the input process over $\mathcal{Y}$, which is a stationary Markov chain with transition probability matrix $\Delta$. Let $Z$ denote the output process over $\mathcal{Z}$, which is often referred to as a \emph{hidden Markov chain}. Assume that $\Delta$ is analytically parameterized by $\theta \in \Omega$, where $\Omega$ is an open, bounded and connected subset of $\mathbb{R}^m$.

Assume that the true parameter of $\Delta$ is $\theta_0$, which is often assumed unknown in a statistical context. For any $l \in \mathbb{N} \cup \{0\}$, we are interested in the limiting probabilistic behavior of the $l$-th derivative of $\log p^{\theta}(Z_1^n)$ with respect to any $\theta \in \Omega$, denoted by $D^l_{\theta} \log p^{\theta}(Z_1^n)$; here $Z_1^n$ is used to denote the sequence of random variables $(Z_1, Z_2, \ldots, Z_n)$, and similar notational convention will be followed in the sequel. We will prove limit theorems for appropriately normalized versions of $D^l_{\theta} \log p^{\theta}(Z_1^n)$, for any fixed $l$ and any $\theta \in \Omega$. Here, we remark that, only for notational convenience, we are treating $\theta$ as a one-dimensional variable throughout this paper.

Consider the following two conditions:
\begin{enumerate}
\item[(I)] $\Pi$ is a strictly positive matrix, and for any $\theta \in \Omega$, $\Delta^{\theta}$ is irreducible and aperiodic;
\item[(II)] for any $\theta \in \Omega$, $\sigma^{(l)}(\theta) \triangleq \lim_{n \to \infty} \sqrt{(\sigma_n^{(l)}(\theta))^2/n} > 0$, where $\sigma_n^{(l)}(\theta)=\sqrt{{\rm Var_{\theta_0}}(D^l_{\theta} \log p^{\theta}(Z_1^n))}$ (the existence of this limit under Condition (I) will be established later).
\end{enumerate}
And we define
$$
L^{(l)}(\theta) \triangleq \lim_{n \to \infty} E_{\theta_0}[D^l_{\theta} \log p^{\theta}(Z_1^n)]/n,
$$
when the limit exists.

The following theorem is an analog of the law of large numbers (LLL).
\begin{thm} \label{HMM-LLL-Theorem}
Assume Condition (I). Then, $L^{(l)}(\theta)$ is well-defined, and for any $\theta \in \Omega$,
$$
\frac{D^l_{\theta} \log p^{\theta}(Z_1^n)}{n} \to L^{(l)}(\theta) \mbox{ with probability $1$}.
$$
\end{thm}
For the case $l=0$, Theorem~\ref{HMM-LLL-Theorem} has already been observed in~\cite{ba66}, where the convergence is used to prove the consistency of the maximum likelihood estimator (MLE) in a hidden Markov model. Note that when $\theta=\theta_0$, we have $L^{(0)}(Z)=-H^{\theta_0}(Z)$, where $H^{\theta_0}(Z)$ denotes the entropy rate of the hidden Markov chain $Z$ at the true parameter $\theta_0$. So, Theorem~\ref{HMM-LLL-Theorem} is a (rather) special case of the celebrated Shannon-McMillan-Breiman theorem, which only assumes the stationarity and ergodicity of $Z$. Entropy rate of a hidden Markov chain is of great importance in many areas in mathematics and physics; in particular, the computation of $H^{\theta_0}(Z)$ is a first step to compute the capacity of a finite-state channel in information theory. Unfortunately, it is notoriously difficult to compute such a fundamental quantity (see~\cite{ep02, mpw11} and references therein). Recently, based on the Shannon-McMillan-Breiman theorem, efficient Monte Carlo methods for approximating $H^{\theta_0}(Z)$ were proposed independently by Arnold and Loeliger~\cite{ar01}, Pfister, Soriaga and Siegel~\cite{pf01}, Sharma and Singh~\cite{sh01}.

We will prove the following central limit theorem (CLT) for $D^l_{\theta} \log p^{\theta}(Z_1^n)$ with an error estimate, which is often referred to as Berry-Esseen bound~\cite{be41, es42} in probability theory. Here, we remark that, in this paper, to avoid notational cumbersomeness, while ensuring its dependence on various variables, we often use $C$ to denote a constant, which may not be the same on each appearance.
\begin{thm} \label{HMM-CLT-Theorem}
Assume Conditions (I) and (II) and consider any given compact subset $\Omega_0 \subset \Omega$. For any $\eps > 0$, there exists $C > 0$ such that for any $n$ and any $\theta \in \Omega_0$,
$$
\sup_x \left|P\left(\frac{D^l_{\theta} \log p^{\theta}(Z_1^n)-n L^{(l)}(\theta)}{\sqrt{n} \sigma^{(l)}(\theta)} < x \right)- G(x) \right| \leq C n^{-1/4+\eps},
$$
where $G(x)=\int_{-\infty}^x (2\pi)^{-1/2} \exp(-y^2/2) dy$.
\end{thm}

For the case $l=1$, Theorem~\ref{HMM-CLT-Theorem} (without the Berry-Esseen bound) has first been shown in~\cite{ba66}, which, together with Theorem~\ref{HMM-LLL-Theorem} for the case $l=2$, can be further used to derive the asymptotic normality of the maximum likelihood estimator (MLE) for a hidden Markov model. This asymptotic normality result is of great importance to the statistical estimation aspects in hidden Markov models, and has been generalized extensively in~\cite{bi96, bi98, do02, do04, le92, me04, ry94, ry97}.

Theorem~\ref{HMM-CLT-Theorem} for the case $l=0$ and $\theta=\theta_0$ (again without the Berry-Esseen bound) has been considered in more probabilistic settings as well: a CLT for $\log p^{\theta}(Z_1^n)$ assuming $Z$ is a Markov chain is first proven in~\cite{yu53}; this result is further generalized to obtain a refinement of the Shannon-McMillan-Breiman theorem in~\cite{ib62} under some mixing assumptions; under somewhat similar conditions, an almost sure invariance principle, a deep result which, among many other applications, implies a CLT, has been established in~\cite{ph75}; the almost sure invariance principle is used to study the asymptotic behavior of the so-called recurrence and waiting times in~\cite{ko98}, where a CLT for $\log p^{\theta}(Z_1^n)$ is embedded in the main results.

In a more information theoretical context, a CLT~\cite{pf03} for $\log p^{\theta}(Z_1^n)$ is derived as a corollary of a CLT for the top Lyapunov exponent of a product of random matrices; a functional CLT is also established in~\cite{ho03}. In essence, both of these two CLTs are proved using effective Martingale approximations of $\log p^{\theta}(Z_1^n)$ (see~\cite{ha80} for this standard technique).

There is also a large body of work (see~\cite{hv04, Haydn09} and references therein) on variants of the CLT for the empirical entropy of some ergodic mappings in the language of ergodic theory, among which, of great relevance to this work are~\cite{hv04, Haydn09}, where CLTs with Berry-Esseen bounds are derived. Here, we remark that there are minor mistakes in the proof of the main results in~\cite{hv04}; it appears that a modified proof, together with stronger assumptions, can only yield weaker results than claimed in~\cite{hv04}.

Note that the error estimate in the CLTs is of great significance in many scenarios, such as characterizing the speed of convergence of the above mentioned Monte Carlo simulation in~\cite{ar01, pf01, sh01} and deriving non-asymptotic coding theorems information theory~\cite{yangenhui} and so on. Among all the previously mentioned related work, only~\cite{hv04, Haydn09} give error estimates for the CLTs. Compared to these two work, where only some mixing conditions are assumed for $Z$, our assumptions are rather strong. On the other hand, our CLT is considerably stronger in the sense that it is essentially for a class of functions including $\log p^{\theta}(Z_1^n)$ and its derivatives with tighter error estimate.

Following Phillip and Stout~\cite{ph75}, we prove the following almost sure invariance principle.
\begin{thm} \label{HMM-ASIP-Theorem}
Assume Conditions (I) and (II). Define a continuous parameter process $\{S(t), t \geq 0\}$ by setting
$$
S(t)=\sum_{n \leq t} D^l_{\theta} \log p^{\theta}(Z_1^n)-n L^{(l)}(\theta).
$$
Then, for any given $\theta \in \Omega$, without changing the distribution of $\{S(t), t \geq 0\}$, we can redefine the process $\{S(t), t \geq 0\}$ on a richer probability space together  with the standard Brownian motion $\{B(t), t \geq 0\}$ such that for any $\eps > 0$,
$$
S(t)-B((\sigma^{(l)}(\theta))^2t)=O(t^{1/3+\eps}) \mbox{ a.s. as $t \to \infty$}.
$$

\end{thm}
As elaborated in~\cite{ph75}, an almost sure invariance principle is a fundamental theorem with many applications, which include, besides a CLT and some large deviation results, a law of iterated logarithm (LIL). The following LIL immediately follows from Theorem~\ref{HMM-ASIP-Theorem}.
\begin{thm} \label{HMM-LIL-Theorem}
Assume Conditions (I) and (II). For any given $\theta \in \Omega$, we have
$$
\limsup_{n \to \infty} \frac{D^l_{\theta} \log p^{\theta}(Z_1^n)-n L^{(l)}(\theta)}{(2n (\sigma^{(l)}(\theta))^2 \log \log n (\sigma^{(l)}(\theta))^2)^{1/2}}=1 \qquad {a.s.}
$$
\end{thm}

Theorem~\ref{HMM-LIL-Theorem} is not completely new: the almost sure invariance principle in~\cite{ph75}, which is established under much weaker conditions, implies Theorem~\ref{HMM-LIL-Theorem} for the case $l=0$. In~\cite{pe66}, it has been shown that with reasonable assumptions, a CLT with a sharp enough error estimation term implies an LIL for i.i.d. sequences of random variables. For possibly dependent sequences of random variables, Petrov's result may not be directly applied to derive an LIL, however the spirit of the proof can be cautiously followed to establish Theorem~\ref{HMM-LIL-Theorem} as an alternative approach (see~\cite{re68}). Using this idea, a law of iterated logarithm (again for the case $l=0$) has also been noted in~\cite{hv04, Haydn09} under some mixing assumptions.

We also prove the following variant of the Chernoff bound (see~\cite{ch52}), giving a sub-exponentially decaying upper bound for the tail probability of $S_n$.
\begin{thm} \label{HMM-Chernoff-Bound-Theorem}
Assume Conditions (I) and (II) and consider any given compact subset $\Omega_0 \subset \Omega$. For any $x > 0$ and any $0 < \eps < 1$, there exist $C > 0$, $0 < \gamma < 1$ such that for any $n$ and any $\theta \in \Omega_0$,
$$
P\left(\frac{D^l_{\theta} \log p^{\theta}(Z_1^n)-n L^{(l)}(\theta)}{n} \geq x \right) \leq C \gamma^{n^{1-\eps}}.
$$
\end{thm}

Let $\theta_n \in \Omega$ be the $n$-th order maximum likelihood estimator (MLE) for the considered hidden Markov model, that is,
$$
\theta_n=\mathrm{argmax}_{\theta \in \Omega} \log p^{\theta}(Z_1^n).
$$
The consistency of the MLE in hidden Markov models have been extensively discussed in statistical contexts (see representative work in~\cite{ba66, le92, bi98}). As one of the principal applications of the limit theorems above, assuming the consistency of the MLE, the following theorem further gives the rate of convergence of the estimators $\theta_n$ to the true parameter $\theta_0$.
\begin{thm} \label{HMM-MLE}
Assume Conditions (I) and (II). Assume that there is a compact subset $\Omega_0 \subset \Omega$ such that $\Omega_0$ contains $\theta_0$ and $L^{(2)}(\theta)$ is non-singular for any $\theta \in \Omega_0$. Then, on the event that ``$\Omega_0$ contains all $\theta_n$'' and ``$\theta_n$ converges to $\theta_0$'', for any $x, \eps > 0$, there exists $C > 0$ such that
$$
P(|\theta_n-\theta_0| \geq x) \leq C n^{-1/4+\eps}.
$$
\end{thm}

\section{Limit Theorems under Exponential Mixing and Forgetting Conditions} \label{Limit-Theorems}

A stationary stochastic process $T=T_{-\infty}^{\infty}$ is said to be {\em $\psi$-mixing} if
$$
\psi(n) \triangleq \sup_{U \in \mathcal{B}(T_{\infty}^{-n}), V \in \mathcal{B}(T_{0}^{\infty}), P(U) > 0, P(V) > 0} |P(V|U)-P(V)|/P(V) \to 0 \mbox{ as } n \to \infty,
$$
where $\mathcal{B}(T_i^j)$ denotes the $\sigma$-field generated by $\{T_k: k=i, i+1, \cdots, j\}$. Let $Z=(Z_n)_{n \in \mathbb{Z}}$ be a stationary $\psi$-mixing sequence of random variables over a finite alphabet $\mathcal{Z}$ satisfying the following property:
\begin{enumerate}
\item[(a)][exponential mixing] There exist $C > 0$, $0 < \lambda < 1$ such that
$$
\psi(n) \leq C \lambda^n
$$
for all $n$.
\end{enumerate}
Let $\mathcal{Z}^*$ be the set of all finite words over $\mathcal{Z}$, and let $f: \mathcal{Z}^* \to \mathbb{R}$ be a function satisfying the following properties:
\begin{enumerate}
\item[(b)] There exist $C', C'' > 0$ such that for all $z_{-n}^0 \in \mathcal{Z}^*$,
$$
C' \leq f(z_0|z_{n-1}^{-1}) \leq C''.
$$
\item[(c)] [exponential forgetting] There exist $C > 0$, $0 < \rho < 1$ such that for any two hidden Markov sequences $z_{-m}^0, \hat{z}_{-\hat{m}}^0$ with $z_{-n}^0=\hat{z}_{-n}^0$ (here $m, \hat{m} \geq n \geq 0$), we have
$$
|f(z_0|z_{-m}^{-1}) - f(\hat{z}_0|\hat{z}_{-\hat{m}}^{-1}) | \leq C \rho^n.
$$
\end{enumerate}
Define
$$
X_i = f(Z_i|Z_1^{i-1})-E[f(Z_i|Z_1^{i-1})],
$$
and
$$
S_n=\sum_{i=1}^n X_i, \qquad \sigma_n^2=Var(S_n).
$$
We will also consider
\begin{enumerate}
\item[(d)]  $\sigma \triangleq \lim_{n \to \infty} \sqrt{\sigma_n^2/n} > 0$ (the existence of this limit under Conditions (a), (b) and (c) will be established in Lemma~\ref{VarianceLemma} and Remark~\ref{existence}).
\end{enumerate}

We will prove the following theorems under Conditions (a), (b), (c) and (d). Not only can these theorems be used to prove the main results in Section~\ref{main-results}, but also they are of interest in their own right. The first theorem is a law of large numbers.
\begin{thm} \label{LLL-Theorem}
Assume Conditions (b) and (c). With probability $1$,
$$
\frac{X_1+X_2+\cdots+X_n}{n} \to 0 \mbox{ as } n \to \infty.
$$
\end{thm}
\noindent We will also prove the following central limit theorem with a Berry-Esseen bound.
\begin{thm} \label{CLT-Theorem}
Assume Conditions (a), (b), (c) and (d). For any $\eps > 0$, there exists $C > 0$ such that for any $n$
$$
\sup_x \left|P(S_n/\sigma_n < x)- G(x) \right| \leq C n^{-1/4+\eps},
$$
where $G(x)=\int_{-\infty}^x (2\pi)^{-1/2} \exp(-y^2/2) dy$.
\end{thm}
\noindent The following theorem is an almost sure invariance principle.
\begin{thm} \label{ASIP-Theorem}
Assume Conditions (a), (b), (c) and (d). Define a continuous parameter process $\{S(t), t \geq 0\}$ by setting
$$
S(t)=\sum_{n \leq t} S_n.
$$
Then, for any given $\theta \in \Omega$, without changing the distribution of $\{S(t), t \geq 0\}$, we can redefine the process $\{S(t), t \geq 0\}$ on a richer probability space together we with the standard Brownian motion $\{B(t), t \geq 0\}$ such that for any $\eps > 0$
$$
S(t)-B(\sigma^2 t)=O(t^{1/3+\eps}) \mbox{ a.s. as $t \to \infty$}.
$$
\end{thm}
\noindent As one of many applications of Theorem~\ref{ASIP-Theorem}, the following law of iterated logarithm immediately follows.
\begin{thm} \label{LIL-Theorem}
Assume Conditions (a), (b), (c) and (d). Then, we have
$$
\limsup_{n \to \infty} \frac{S_n}{(2n \sigma^2 \log \log n \sigma^2)^{1/2}}=1 \qquad {a.s.}
$$
\end{thm}
\noindent We also prove the following variant of the Chernoff bound (see~\cite{ch52}), giving a sub-exponentially decaying upper bound for the tail probability of $S_n$.
\begin{thm} \label{Chernoff-Bound-Theorem}
Assume Conditions (a), (b) and (c). For any $x > 0$ and any $0 < \eps < 1$, there exist $C > 0$, $0 < \gamma < 1$ such that for any $n$,
$$
P(S_n/n \geq x) \leq C \gamma^{n^{1-\eps}}.
$$
\end{thm}

\section{Proofs of the Theorems in Section~\ref{Limit-Theorems}}

\subsection{Key Lemmas} \label{Key-Lemmas}

From now on, we rewrite $f(z_{j}|z_{i}^{j-1})-E[f(Z_{j}|Z_{i}^{j-1})]$ as $g(z_i^j)$ for notational simplicity.

The following lemma shows that for a fixed $j > 0$, $E[X_i X_{i+j}]$ exponentially converges as $i \to \infty$, and for any $i < j$, $E[X_i X_j]$ exponentially decays in $j-i$.
\begin{lem}  \label{two}
Assume that Conditions (a), (b) and (c).
\begin{enumerate}
\item There exist $C > 0$, $0 < \rho < 1$ (here $\rho$ is as in Condition (c)) such that for all $i, j \geq 0$,
$$
|E[X_{i+1}X_{i+1+j}]-E[X_{i}X_{i+j}]| \leq C \rho^{i}.
$$
\item There exist $C > 0$, $0 < \theta < 1$ such that for any positive $i < j$,
$$
|E[X_i X_j]| \leq C \theta^{j-i}.
$$
\end{enumerate}
\end{lem}

\begin{proof}
1. Simple computations lead to
\begin{eqnarray} \notag
\hspace{-2cm} E[X_{i+1}X_{i+1+j}]-E[X_iX_{i+j}]&=&\sum_{z_1^{i+1+j}} p(z_1^{i+1+j}) g(z_1^{i+1+j}) g(z_1^{i+1})-\sum_{z_1^{i+j}} p(z_1^{i+j}) g(z_1^{i+j}) g(z_1^{i})\\
\notag &=&\sum_{z_{-i-j}^{0}} p(z_{-i-j}^{0}) g(z_{-i-j}^0) g(z_{-i-j}^{-j})
-\sum_{z_{-i-j+1}^{0}} p(z_{-i-j+1}^{0}) g(z_{-i-j+1}^0) g(z_{-i-j+1}^{-j})\\
\notag &=&\sum_{z_{-i-j}^{0}} p(z_{-i-j}^{0}) (g(z_{-i-j}^0) g(z_{-i-j}^{-j})
-g(z_{-i-j+1}^0) g(z_{-i-j+1}^{-j}))\\
\notag &=&\sum_{z_{-i-j}^{0}} p(z_{-i-j}^{0}) g(z_{-i-j}^0) (g(z_{-i-j}^{-j})-g(z_{-i-j+1}^{-j}))\\
       & & +\sum_{z_{-i-j}^{0}} p(z_{-i-j}^{0}) (g(z_{-i-j}^0)-g(z_{-i-j+1}^0)) g(z_{-i-j+1}^{-j}).
\end{eqnarray}
By Condition (b), $f(z_{0}|z_{-i}^{-1})$ and $E[f(Z_0|Z_{-i}^{-1})]$ are all bounded from above and below uniformly in $i$. It then follows from this fact and Condition (c) that there exist $C > 0$, $0 < \rho < 1$ such that
$$
|E[X_{i+1}X_{i+1+j}]-E[X_i X_{i+j}]| \leq C \rho^i.
$$
Part $1$ of the lemma then immediately follows.

2. Let $l=\lfloor i+j \rfloor/2$. By Conditions (a) and (c), there exist $0 < \rho, \lambda < 1$ such that
\begin{eqnarray}
\notag E[X_i X_j]&=&\sum_{z_1^j} p(z_1^j) g(z_1^i) g(z_1^j)\\
\notag &=&\sum_{z_1^j} p(z_1^j) g(z_1^j)(g(z_l^j)+O(\rho^{j-l}))\\
\notag &=&\sum_{z_1^i, z_l^j} p(z_1^i) g(z_1^i) p(z_l^j|z_1^i) g(z_l^j)+O(\rho^{j-l})\\
\notag &=&\sum_{z_1^i, z_l^j} p(z_1^i) g(z_1^i) (p(z_l^j)+O(\lambda^{l-i})p(z_l^j)) g(z_l^j)+O(\rho^{j-l})\\
\notag &=&\sum_{z_1^i, z_l^j} p(z_1^i) g(z_1^i) p(z_l^j) g(z_l^j) +\sum_{z_1^i, z_l^j} p(z_1^i) g(z_1^i) O(\lambda^{l-i})  p(z_l^j) g(z_l^j)+O(\rho^{j-l})\\
\notag &=&0+O(\lambda^{l-i})+O(\rho^{j-l}).
\end{eqnarray}
Notice that the constants in $O(\lambda^{l-i})$, $O(\rho^{j-l})$ above do not depend on $z_1^j$. Part $2$ then immediately follows .
\end{proof}

\begin{rem} \label{a_j}
By Part $1$ of Lemma~\ref{two}, for any fixed $j$, the sequence $E[X_i X_{i+j}]$, $i=1, 2, \cdots$, is a Cauchy sequence that exponentially converges. For any fixed $j$, let $a_j=\lim_{i \to \infty} E[X_i X_{i+j}]$. Then by Part $2$, $|a_j|$ exponentially decays as $j \to \infty$; consequently, we deduce (for later use) that $a_0+2\sum_{j=1}^{\infty} a_j$ converges.
\end{rem}

\begin{lem} \label{VarianceLemma}
Assume Conditions (a), (b) and (c). For any $0 < \eps_0 < 1$, there exists $C > 0$ such that for any $m$ and $n$,
$$
\left|\frac{E[(S_{n+m}-S_m)^2]}{n}-(a_0+2\sum_{j=1}^{\infty} a_j) \right| \leq C n^{-\eps_0};
$$
here, recall that, as defined in Remark~\ref{a_j}, $a_j=\lim_{i \to \infty} E[X_i X_{i+j}]$.
\end{lem}

\begin{proof}
Letting $\beta=n^{-\eps_0}$ for a fixed $0 < \eps_0 < 1$, we then have
$$
\hspace{-1cm} \frac{E[(S_{n+m}-S_m)^2]}{n}=\frac{E[(\sum_{i=m+1}^{n+m} X_i)^2]}{n}=\frac{\sum_{m+1 \leq i, \;\; i+j \leq n+m} (\sum_{j=0}+2\sum_{0 < j \leq \beta n} + 2\sum_{j > \beta n})E[X_i X_{i+j}]}{n}.
$$
By Part $1$ of Lemma~\ref{two} and Remark~\ref{a_j}, for any $j > 0$, $E[X_i X_{i+j}]-a_j=O(\rho^i)$ for some $0 < \rho < 1$. It then follows that for $0 \leq j \leq \beta n$,
$$
\sum_{m+1 \leq i, \;\; i+j \leq n+m} E[X_i X_{i+j}]=(n-j)a_j+O(1);
$$
here the constant in $O(1)$ does not depend on $j$. Also, by Part $2$ of Lemma~\ref{two} and Remark~\ref{a_j}, there exists $0 < \theta < 1$ such that for all $j > \beta n$, $E[X_i X_{i+j}]=O(\theta^{\beta n})$, and thus $a_j=O(\theta^{\beta n})$. Continuing the computation, we have
$$
\hspace{-1cm} \frac{E[(S_{n+m}-S_m)^2]}{n}=\frac{(na_0+O(1))+(2(n-1)a_1+O(1))+\cdots+(2(n-\beta n)a_{\beta n}+O(1))}{n} + \frac{O(n^2 \theta^{\beta n})}{n}
$$
$$
=a_0+2a_1+\cdots+2a_{\beta n}-2\frac{a_1+2 a_2+ \cdots+ \beta n a_{\beta n}}{n}+\beta O(1)+ O(n \theta^{\beta n}).
$$
The lemma then immediately follows if we let $n$ go to infinity.
\end{proof}

\begin{rem} \label{existence}
Choosing $m$ in Lemma~\ref{VarianceLemma} to be $0$, we deduce that $\lim_{n \to \infty} \sigma_n^2/n$ exists and is equal to $\sigma^2=a_0+2\sum_{j=1}^{\infty} a_j$.
\end{rem}

\begin{lem} \label{growth-of-even-moments}
For any $l \in \mathbb{N}$, there exists $C > 0$ such that for all $m$ and $n$,
$$
E[(S_{n+m}-S_m)^{2 l}] \leq C n^l.
$$
\end{lem}

\begin{proof}
By Condition (c) and the stationarity of $Z$, we observe that for any $m$, $n$,
$$
E[(S_{n+m}-S_{m})^{2 l}]=E[(\sum_{i=m+1}^{n+m} g(Z_1^i))^{2 l}]=E[(\sum_{i=m+1}^{n+m} (g(Z_{m+1}^i)+O(\rho^{i-m-1})))^{2 l}]
$$
\begin{equation}
=E[S_n^{2 l}]+O(E[|S_n|^{2 l-1}])+O(E[S_n^{2 l -2}])+ \cdots+ O(1).
\end{equation}
Notice that for any $j$,
$$
E[|S_n|^{2 l-1}] \leq E[S_n^{2 l}]^{1/2} E[S_n^{2 l-2}]^{1/2}.
$$
So, in order to prove the lemma, it suffices to prove that for any $l \in \mathbb{N}$, there exists $C_1 > 0$ such that
$$
E[S_n^{2 l}] = E[(X_1+X_2+\cdots+X_n)^{2 l}] \leq C_1 n^l.
$$
Now, for any $l \in \mathbb{N}$, consider the term $X_{i_1}^{l_1} X_{i_2}^{l_2} \cdots X_{i_k}^{l_k}$, where $1 \leq i_1 < i_2 < \cdots < i_k \leq n$ and $l_j$'s are all strictly positive satisfying $l_1+l_2+\cdots+l_k \leq 2l$. Let $v \triangleq v(i_1, i_2, \ldots, i_k)$ be the smallest index such that for all $j=1, 2, \ldots, k-1$,
\begin{equation} \label{v-defined}
i_{v+1}-i_v \geq i_{j+1}-i_j.
\end{equation}
Now, for any $v+1 \leq u \leq k$, recalling that 
$$
X_{i_u}=f(Z_{i_u}|Z_{1}^{i_u-1})-E[f(Z_{i_u}|Z_{1}^{i_u-1})],
$$
we define 
$$
\tilde{X}_{i_u}=f(Z_{i_u}|Z_{(i_v+i_{v+1})/2}^{i_u-1})-E[f(Z_{i_u}|Z_{(i_v+i_{v+1})/2}^{i_u-1})].
$$
Applying Condition (c), we have for some $0 < \rho < 1$
$$
X_{i_u}-\tilde{X}_{i_u}=O(\rho^{(i_{v+1}-i_v)/2}).
$$
We then have the following decomposition:
\begin{align}
\nonumber E[X_{i_1}^{l_1} \cdots X_{i_v}^{l_v} X_{i_{v+1}}^{l_{v+1}} \cdots X_{i_k}^{l_k}] &= E[X_{i_1}^{l_1} \cdots X_{i_v}^{l_v} (\tilde{X}_{i_{v+1}}+O(\rho^{(i_{v+1}-i_v)/2}))^{l_{v+1}} \cdots (X_{i_k}+O(\rho^{(i_{v+1}-i_v)/2}))^{l_k}]\\
\nonumber &= E[X_{i_1}^{l_1} \cdots X_{i_v}^{l_v} \tilde{X}_{i_{v+1}}^{l_{v+1}} \cdots \tilde{X}_{i_k}^{l_k}]+r^{(1)}[X_{i_1}^{l_1} X_{i_2}^{l_2} \cdots X_{i_k}^{l_k}]\\
\nonumber &= E[X_{i_1}^{l_1} \cdots X_{i_v}^{l_v}] E[\tilde{X}_{i_{v+1}}^{l_{v+1}} \cdots \tilde{X}_{i_k}^{l_k}]+r^{(2)}[X_{i_1}^{l_1} X_{i_2}^{l_2} \cdots X_{i_k}^{l_k}]\\
\nonumber &= E[X_{i_1}^{l_1} \cdots X_{i_v}^{l_v}] E[(X_{i_{v+1}}+O(\rho^{(i_{v+1}-i_v)/2}))^{l_{v+1}} \cdots (X_{i_k}+O(\rho^{(i_{v+1}-i_v)/2}))^{l_k}]\\
\nonumber &\phantom{abcd} +r^{(2)}[X_{i_1}^{l_1} \cdots X_{i_v}^{l_v} X_{i_{v+1}}^{l_{v+1}} \cdots X_{i_k}^{l_k}]\\
\nonumber &= E[X_{i_1}^{l_1} X_{i_2}^{l_2} \cdots X_{i_v}^{l_v}] E[X_{i_{v+1}}^{l_{v+1}} X_{i_{v+2}}^{l_{v+2}} \cdots X_{i_k}^{l_k}]+r[X_{i_1}^{l_1} X_{i_2}^{l_2} \cdots X_{i_k}^{l_k}],
\end{align}
where $r^{(1)}[X_{i_1}^{l_1} X_{i_2}^{l_2} \cdots X_{i_k}^{l_k}], r^{(2)}[X_{i_1}^{l_1} X_{i_2}^{l_2} \cdots X_{i_k}^{l_k}]$ are some intermediate terms produced during the decomposition and $r[X_{i_1}^{l_1} X_{i_2}^{l_2} \cdots X_{i_k}^{l_k}]$ is the residual term resulted from the decomposition. Using (\ref{v-defined}) and Conditions (a), (b), (c), we can verify that for some $0 < \theta < 1$
\begin{equation} \label{residual-term-estimation}
\sum_{l_j > 0, \; \sum_{j} l_j \leq 2 l} r[X_{i_1}^{l_1} X_{i_2}^{l_2} \cdots X_{i_k}^{l_k}]=\sum_{v=1}^{n-1} \sum_j O(\theta^j ((2l-2)j)^{2l-2}) = O(n).
\end{equation}
Note that the above decomposition can be recursively applied to $E[X_{i_1}^{l_1} X_{i_2}^{l_2} \cdots X_{i_k}^{l_v}]$ and $E[X_{i_{v+1}}^{l_{v+1}} X_{i_{v+2}}^{l_{v+2}} \cdots X_{i_k}^{l_k}]$. It then follows that $E[X_{i_1}^{l_1} X_{i_2}^{l_2} \cdots X_{i_k}^{l_k}]$ can be decomposed into a sum of at most $2^{2l}$ terms, each of which taking the following form
$$
E[X_{i'_1}^{l'_1}] E[X_{i'_2}^{l'_2}] \cdots E[X_{i'_{k_1}}^{l'_{k_1}}] r_{i^*_1} r_{i^*_2} \cdots r_{i^*_{k_2}},
$$
where each $l'_j \geq 2$, $l'_1+l'_2+\cdots+l'_{k_1}+2 k_2 \leq 2l$ and $r_{i^*_1}, r_{i^*_2}, \cdots ,r_{i^*_{k_2}}$ are the residual terms resulted from the recursive decomposition. Then, similarly as in deriving (\ref{residual-term-estimation}), one checks that $E[S_n^{2 l}]$ can be written as a sum of at most $2^{2l}$ terms, each of which is upper bounded by
$$
(\sum E[|X_{i'_1}|^{l'_1}] E[|X_{i'_2}|^{l'_2}] \cdots E[|X_{i'_{k_1}}|^{l'_{k_1}}]) \underbrace{O(n) \cdots O(n)}_{k_2},
$$
where
\begin{equation} \label{l-prime-conditions}
l'_j \geq 2, \qquad l'_1+l'_2+\cdots+l'_{k_1}+2 k_2 \leq 2l,
\end{equation} 
and the summation is over all possible $X_{i'_1}^{l'_1} X_{i'_2}^{l'_2} \cdots X_{i'_{k_1}}^{l'_{k_1}}$ satisfying (\ref{l-prime-conditions}), which can be estimated by
$$
\sum E[|X_{i'_1}|^{l'_1}] E[|X_{i'_2}|^{l'_2}] \cdots E[|X_{i'_{k_1}}|^{l'_{k_1}}] = O(n^{l-k_2}).
$$
It then follows that 
$$
E[S_n^{2 l}]=O(n^{l-k_2}) O(n^{k_2}) = O(n^l).
$$
We then have established the lemma.
\end{proof}

\begin{lem} \label{growth-of-odd-moments}
For any $l \in \mathbb{N}$, there exists $C > 0$ such that for all $m$ and $n$,
$$
E[|S_{n+m}-S_m|^{2 l-1}] \leq C n^{l-1/2}.
$$
\end{lem}

\begin{proof}
The lemma immediately follows from Lemma~\ref{growth-of-even-moments} and the fact that for any $m, n$,
$$
E[|S_{n+m}-S_m|^{2 l-1}] \leq E[(S_{n+m}-S_{m})^{2 l}]^{1/2} E[(S_{n+m}-S_m)^{2 l-2}]^{1/2}.
$$
\end{proof}

\subsection{Proof of Theorem~\ref{LLL-Theorem}}
It follows from Condition (c) that there exists $0 < \rho < 1$ such that for any $j < i$,
$$
|f(Z_i|Z_j^{i-1})-f(Z_i|Z_{j-1}^{i-1})| = O(\rho^{i-j}),
$$
which implies that $f(Z_i|Z_{-\infty}^{i-1}) \triangleq \lim_{j \to -\infty} f(Z_i|Z_j^{i-1})$ exists, and
$$
|f(Z_i|Z_j^{i-1})-f(Z_i|Z_{-\infty}^{i-1})|=O(\rho^{i-j}),
$$
and furthermore
$$
|E[f(Z_i|Z_j^{i-1})]-E[f(Z_i|Z_{-\infty}^{i-1})]|=O(\rho^{i-j}).
$$
We then have
$$
\frac{\sum_{i=1}^n X_i}{n}=\sum_{i=1}^n \frac{f(Z_i|Z_1^{i-1})-E[f(Z_i|Z_1^{i-1})]}{n}=\sum_{i=1}^n \frac{f(Z_i|Z_{-\infty}^{i-1})-E[f(Z_i|Z_{-\infty}^{i-1})]+O(\rho^i)}{n}.
$$
Here, we remark that the constants in all the above $O$-terms are independent of $i, j$. Note that the sequence $f(Z_i|Z_{-\infty}^{i-1})-E[f(Z_i|Z_{-\infty}^{i-1})]$ is stationary and ergodic. Applying the Birkhoff ergodic theorem, and using the fact that $\sum_{i=1}^n \rho^i/n \to 0$ as $n \to \infty$, we then establish the theorem.

\subsection{Proof of Theorem~\ref{CLT-Theorem}}

For any fixed $0 < \beta < \alpha < 1$, we consecutively partition the partial sum $S_n$ into blocks $\eta_1, \zeta_1, \eta_2, \zeta_2, \ldots$ such that each $\eta_i$ is of length $p=p(n) \triangleq n^{\beta}$ and each $\zeta_i$ is of length $q=q(n) \triangleq n^{\alpha}$. In other words, for any feasible $i$,
$$
\eta_i = X_{(i-1)q+(i-1)p+1}+\cdots+X_{i q+(i-1)p},
$$
and
$$
\zeta_i = X_{i q+ (i-1)p +1}+ \cdots+ X_{i q+i p}.
$$
Then, $S_n$ can be rewritten as a sum of $\eta$-``blocks'' and $\zeta$-``blocks''
$$
S_n=S^*_n + S'_n := \sum_{i=1}^k \eta_i + \sum_{i=1}^k \zeta_i,
$$
where $k=k(n) \triangleq n/(n^{\alpha}+n^{\beta})$. The above so called Bernstein blocking method~\cite{bi26} is a standard technique for proving limit theorems for a variety of mixing sequences. Roughly speaking, the partial sum $S_n$ is partitioned into ``short blocks'' $\eta_1, \eta_2, \cdots, \eta_k$ and ``long blocks'' $\zeta_1, \zeta_2, \cdots, \zeta_k$. Under certain mixing conditions, all long blocks are ``weakly dependent'' on each other, while all short blocks are ``negligible'' in some sense.

Now, we will ``truncate'' $\zeta_i$'s to obtain $\hat{\zeta}_i$'s. In more detail, recall that for any $j$ with $i q+ (i-1)p+1 \leq j \leq i q + i p$, we have
$$
X_j=f(Z_j|Z_1^{j-1})-E[f(Z_j|Z_1^{j-1})];
$$
we then define
$$
\hat{X}_j=f(Z_j|Z_{(i-1)p+(i-1)q+\lfloor q/2 \rfloor+1}^{j-1})-E[f(Z_j|Z_{(i-1)p+(i-1)q+\lfloor q/2 \rfloor+1}^{i-1})].
$$
Applying Condition (c), we derive that
\begin{equation} \label{x-and-xhat}
X_j-\hat{X}_j=O(\rho^{q(n)/2}).
\end{equation}
We then define,
$$
\hat{\zeta}_i = \hat{X}_{i q+ (i-1)p +1}+ \cdots+ \hat{X}_{i q+i p},
$$
and
$$
S'_n=\sum_{i=1}^k \zeta_i, \qquad \hat{\sigma}'_n=\sqrt{{\rm Var}(\hat{S}'_n)}.
$$

With lemmas in Section~\ref{Key-Lemmas} established, the remainder of the proof of Theorem~\ref{CLT-Theorem} becomes more or less standard, which can be roughly outlined as follows:
\begin{enumerate}
\item We first show $E[\exp(i t \hat{S}'_n/\sigma_n)]$ and $\prod_{j=1}^k E[\exp (it \hat{\zeta}_j/\sigma_n)$ are ``close'' (see Lemma~\ref{phi2phi1}).
\item Then by the standard Esseen's Lemma, we show $P(\hat{S}'_n/\sigma_n < x)$ and $G(x)$ are ``close'' (see Lemma~\ref{SnPrime}).
\item Finally, since $S^*_n$ are ``negligible'', we conclude, in the proof of Theorem~\ref{CLT-Theorem}, that $P(S_n/\sigma_n < x)$ and $P(\hat{S}'_n/\sigma_n < x)$ are ``close'', and thus $P(S_n/\sigma_n < x)$ and $G(x)$ are ``close''.
\end{enumerate}

Before proceeding, we first remind the reader the classical Esseen\rq{}s inequality (see, e.g., Lemma $5.1$ on Page $147$ of~\cite{pe95}).
\begin{lem}[Esseen's Inequality] \label{Esseen-Inequality}
Let $\bar{\zeta}_1, \bar{\zeta}_2, \cdots, \bar{\zeta}_n$ be independent random variables with $E[\bar{\zeta}_j]=0$, $E[|\bar{\zeta}_j|^3] < \infty$, $j=1, 2, \cdots, n$. Let
$$
\bar{\sigma}_n^2=\sum_{j=1}^n E[\bar{\zeta}_j^2], \quad L_n=\bar{\sigma}_n^{-3} \sum_{j=1}^n E[|\bar{\zeta}_j|^3],
$$
and let $\bar{F}_n(x), \phi_{\bar{F}_n}(t)$ be the distribution, characteristic functions of the random variable $\sum_{j=1}^n \bar{\zeta}_j/\bar{\sigma}_n$, respectively. Then
\begin{equation}
|\phi_{\bar{F}_n}(t)-e^{-t^2/2}| \leq 16 L_n |t|^3 e^{-t^2/3}
\end{equation}
for $|t| \leq 1/(4 L_n)$.
\end{lem}

The following lemma is a version of Esseen's lemma, which gives an upper bound on the difference between two distribution functions using the difference between the two corresponding characteristic functions. We refer to page $314$ of~\cite{st74} for a standard proof.
\begin{lem}[Esseen's Lemma] \label{Esseen}
Let $F(x)$ and $G(x)$ be distribution functions with characteristic functions $\phi_F(t)$ and $\phi_G(t)$, respectively. Suppose that the distributions corresponding to $F(x)$ and $G(x)$ each has mean $0$, and $G(x)$ is differentiable and for any $x$, $|G'(x)| \leq M$ for some $M > 0$. Then
$$
\sup_x |F(x)-G(x)| \leq \frac{1}{\pi} \int_{-T}^{T} \left|\frac{\phi_F(t)-\phi_G(t)}{t}\right| dt+\frac{24M}{\pi T}
$$
for every $T > 0$.
\end{lem}

We will need the following lemma.
\begin{lem} \label{phi2phi1}
There exist $C > 0$, $0 < \rho_1 < 1$ such that for all $n$ and $|t| \leq n^{1/2}$,
$$
|E[\exp(i t \hat{S}'_n/\hat{\sigma}_n\rq{})]-\prod_{j=1}^k E[\exp (it \hat{\zeta}_j/\hat{\sigma}_n\rq{})]| \leq C \rho_1^{q(n)}.
$$
\end{lem}

\begin{proof}
Let $l=(k-1)p+(k-1)q+\lfloor q/2 \rfloor+1$. By Condition (a), there exists $0 < \lambda < 1$ such that
\begin{eqnarray}
\notag \hspace{-2cm} E[\exp (i t \sum_{j=1}^{k} \hat{\zeta}_j/\hat{\sigma}_n\rq{})]&=&E[\exp(i t \sum_{j=1}^{k-1} \hat{\zeta}_j/\hat{\sigma}_n\rq{}) \exp (i t \hat{\zeta}_k/\hat{\sigma}_n\rq{})]\\
\notag &=&E[\exp (i t \sum_{j=1}^{k-1} \hat{\zeta}_j/\hat{\sigma}_n\rq{}) \exp (i t \sum_{i=kq+(k-1)p+1}^{kq+kp} g(z_l^i)/\hat{\sigma}_n\rq{})]\\
\notag &=&E[\exp (i t \sum_{j=1}^{k-1} \hat{\zeta}_j/\hat{\sigma}_n\rq{} )]E[\exp (i t \sum_{i=kq+(k-1)p+1}^{kq+kp} g(z_l^i)/\hat{\sigma}_n\rq{})]+O(\lambda^{q(n)/2})\\
\notag &=&E[\exp (i t \sum_{j=1}^{k-1} \hat{\zeta}_j/\hat{\sigma}_n\rq{})] E[\exp (i t \hat{\zeta}_k/\hat{\sigma}_n\rq{})]+O(\lambda^{q(n)/2}),
\end{eqnarray}
where, again, $f(z_{j}|z_{i}^{j-1})-E[f(Z_{j}|Z_{i}^{j-1})]$ is rewritten as $g(z_i^j)$. Noticing that $|E[\exp (it \hat{\zeta}_j/\hat{\sigma}_n\rq{})]| \leq 1$ and applying an inductive argument, we conclude that
$$
E[\exp (i t \hat{S}'_n/\hat{\sigma}_n\rq{})]=E[\exp (i t \sum_{j=1}^{k} \hat{\zeta}_j/\hat{\sigma}_n\rq{})]=\prod_{j=1}^k E[\exp (it \hat{\zeta}_j/\hat{\sigma}_n\rq{})]| +O(k \lambda^{q(n)/2}),
$$
which immediately implies the lemma.
\end{proof}

Now, applying Lemma~\ref{Esseen}, we can derive the following lemma.
\begin{lem} \label{SnPrime}
There exists $C > 0$ such that for all $n$
$$
\sup_x \left|P(\hat{S}'_n/\hat{\sigma}'_n < x)-G(x) \right| \leq C n^{-1/2+\alpha/2}.
$$
\end{lem}

\begin{proof}

Note that all $\hat{\zeta}_j$\rq{}s have the same distribution. So, Lemma~\ref{phi2phi1} in fact implies that
\begin{equation} \label{approximated-by-truncated}
|E[\exp(i t \hat{S}'_n/\hat{\sigma}_n\rq{})]-(E[\exp (it \hat{\zeta}_1/\hat{\sigma}_n\rq{})])^k| = O(\rho_1^{q(n)}),
\end{equation}
for some $0 < \rho_1 < 1$. Consider a sequence of i.i.d. random variables $\bar{\zeta}_j$, $j=1, 2, \cdots,k$, each of which is distributed according to $\hat{\zeta}_1$. It then follows from (\ref{approximated-by-truncated}) that
\begin{equation} \label{approximated-by-independent}
|E[\exp(i t \hat{S}'_n/\hat{\sigma}_n\rq{})]-(E[\exp (it \bar{\zeta}_1/\hat{\sigma}_n\rq{})])^k| = O(\rho_1^{q(n)}).
\end{equation}
Now, let
\begin{equation} \label{bar-variance}
\bar{\sigma}_n^2=\sum_{j=1}^k E[\bar{\zeta}_j^2].
\end{equation}
It follows from Condition (a) that for some $0 < \lambda < 1$
$$
(\hat{\sigma}'_n)^2-\bar{\sigma}_n^2=O(k^2 E^2[|\hat{\zeta}_1|] \lambda^{q(n)}),
$$
which implies that
\begin{equation} \label{something-tricky}
(E[\exp (it \bar{\zeta}_1/\hat{\sigma}'_n)])^k-(E[\exp (it \bar{\zeta}_1/\bar{\sigma}_n)])^k=O(\rho_2^{q(n)}),
\end{equation}
for some $0 < \rho_2 < 1$. Therefore, combining (\ref{approximated-by-independent}) and (\ref{something-tricky}), we deduce that
\begin{equation} \label{approximated-by-normalized-independent}
|E[\exp(i t \hat{S}'_n/\hat{\sigma}_n\rq{})]-(E[\exp (it \bar{\zeta}_1/\bar{\sigma}_n)])^k| = O(\rho_3^{q(n)}),
\end{equation}
for some $0 < \rho_3 < 1$. So, in the sense of (\ref{approximated-by-normalized-independent}), we can approximate $\hat{S}'_n/\hat{\sigma}_n'$ using the sum of i.i.d random variables $\bar{\zeta}_j/\bar{\sigma}_n$, $j=1, 2, \cdots, k$, each of which is distributed according to $\hat{\zeta}_1/\bar{\sigma}_n$. Applying Lemma~\ref{Esseen-Inequality} to the i.i.d. sequence $\bar{\zeta}_j/\bar{\sigma}_n$, we deduce that for $|t| \leq 1/(4 L_n)$,
\begin{equation} \label{better-than-Reznik}
|(E[\exp (it \bar{\zeta}_1/\bar{\sigma}_n)])^k-e^{-t^2/2}| \leq 16 L_n |t|^3 e^{-t^2/3}
\end{equation}
where
$$
L_n = \sum_{j=1}^k E[|\bar{\zeta}_j|^3]/\bar{\sigma}_n^{3}=k E[|\hat{\zeta}_1|^3]/\bar{\sigma}_n^{3} .
$$
Note that, by (\ref{bar-variance}) and Lemma~\ref{VarianceLemma}, we have
$$
\bar{\sigma}_n^3=\Theta(n^{3/2}).
$$
Furthermore, by (\ref{x-and-xhat}) and Lemma~\ref{growth-of-odd-moments}, we have
$$
k E[|\hat{\zeta}_1|^3]= k O(p(n)^{3/2}) = O(n^{1+\alpha/2}).
$$
It then follows that there exists $C_1 > 0$ such that for all $n$,
\begin{equation} \label{order-of-Ln}
L_n \leq C_1 n^{-1/2+\alpha/2}.
\end{equation}

From now on, let $\phi_{\bar{F}_n}(t), \phi_{\hat{F}'_n}(t)$ be the characteristic functions of the random variable $\sum_{j=1}^k \bar{\zeta}_j/\bar{\sigma}_n, \hat{S}'_n/\hat{\sigma}'_n$, respectively. Then, by Lemma~\ref{Esseen}, we have
$$
\sup_x |P(\hat{S}'_n/\hat{\sigma}'_n < x)-G(x)| \leq \frac{1}{\pi} \int_{-T}^{T} \left|\frac{\phi_{\hat{F}'_n}(t)-\phi_G(t)}{t}\right| dt+\frac{24M}{\pi T}
$$
for every $T > 0$. It then follows that for any $T > 0$
$$
\sup_x |P(\hat{S}'_n/\hat{\sigma}'_n < x)-G(x)| \leq \frac{1}{\pi} \int_{-T}^{T} \left|\frac{\phi_{\hat{F}'_n}(t)-\phi_{\bar{F}_n}(t)}{t}\right| dt+\frac{1}{\pi} \int_{-T}^{T} \left|\frac{\phi_{\bar{F}_n}(t)-\phi_G(t)}{t}\right| dt+\frac{24M}{\pi T}
$$
$$
\hspace{-1cm} \leq \frac{1}{\pi} \int_{|t| \leq n^{-1/2}} \left|\frac{\phi_{\hat{F}'_n}(t)-\phi_{\bar{F}_n}(t)}{t}\right| dt+\frac{1}{\pi} \int_{n^{-1/2} \leq |t| \leq T} \left|\frac{\phi_{\hat{F}'_n}(t)-\phi_{\bar{F}_n}(t)}{t}\right| dt+\frac{1}{\pi} \int_{-T}^{T} \left|\frac{\phi_{\bar{F}_n}(t)-\phi_G(t)}{t}\right| dt+\frac{24M}{\pi T}.
$$
Note that there exists $C_2 > 0$ such that for all $t$,
\begin{equation} \label{by-Reznik}
|\phi_{\hat{F}'_n}(t)-\phi_{\bar{F}_n}(t)| \leq C_2 t.
\end{equation}
Now, setting $T=1/(4 C_1 n^{-1/2+\alpha/2})$ and applying (\ref{better-than-Reznik}), (\ref{by-Reznik}) and (\ref{approximated-by-normalized-independent}), we then have
$$
\hspace{-2cm} \sup_x |P(\hat{S}'_n/\hat{\sigma}'_n < x)-G(x)| \leq \frac{2 C_2}{\pi} n^{-1/2}+\frac{2-\alpha}{\pi} \log n \rho_3^{q(n)}-\frac{2}{\pi} \log (4 C_1) \rho_3^{q(n)} + \frac{16 L_n}{\pi} \int_{-\infty}^{\infty} t^2 e^{-t^2/3} dt + \frac{96 M C_1}{\pi} n^{-1/2+\alpha/2},
$$
which immediately implies the lemma.

\end{proof}

We are now ready to prove Theorem~\ref{CLT-Theorem}. The key point is $P(S_n/\sigma_n \leq x)$ is close to $P(\hat{S}'_n/\hat{\sigma}'_n \leq x)$.

\begin{proof}[Proof of Theorem~\ref{CLT-Theorem}]

Applying Conditions (a), (c) and Lemma~\ref{VarianceLemma}, we deduce that for any small $\eps_0 > 0$,
$$
\sigma_n^2=\sigma n + O(n^{\eps_0}),
$$
and
\begin{align*}
(\hat{\sigma}'_n)^2 & =E[(\sum_{i=1}^k \hat{\zeta_i})^2]\\
&=\sum_{i=1}^k E[\hat{\zeta_i}^2] + 2\sum_{i < j} E[\hat{\zeta_i} \hat{\zeta_j}]\\
&=k E[\hat{\zeta_1}^2] + 2\sum_{i < j} E[\hat{\zeta_i} \hat{\zeta_j}]=k E[(\zeta_1+O(n^{\alpha}\rho^{q(n)/2}))^2] + 2\sum_{i < j} E[\hat{\zeta_i} \hat{\zeta_j}]\\
&=k E[\zeta_1^2]+O(k n^{2\alpha} \rho^{q(n)})+O(k n^{\alpha} E[|\zeta_1|] \rho^{q(n)})+ O(k^2 E^2[|\hat{\zeta}_1|] \lambda^{q(n)/2})\\
&=k (\sigma n^{\alpha}+O(n^{\eps_0}))+ O(k n^{2\alpha} \rho^{q(n)})+O(k n^{\alpha} E[|\zeta_1|] \rho^{q(n)})+ O(k^2 E^2[|\hat{\zeta}_1|] \lambda^{q(n)/2})\\
&=\sigma \frac{n}{n^{\alpha}+n^{\beta}} n^{\alpha}+ \frac{n}{n^{\alpha}+n^{\beta}} O(n^{\eps_0})+ O(k n^{2\alpha}\rho^{q(n)})+O(k n^{\alpha} E[|\zeta_1|] \rho^{q(n)})+ O(k^2 E^2[|\hat{\zeta}_1|] \lambda^{q(n)/2}).
\end{align*}
It then follows that
$$
\sigma_n^2-(\hat{\sigma}'_n)^2=O(n^{1-\alpha+\beta}).
$$
Next, applying Condition (b) and Lemma~\ref{VarianceLemma}, we have, through simple computations, that
$$
\qquad \hat{S}'_n=O \left(\frac{n^{\alpha}}{n^{\alpha}+n^{\beta}} n \right) = O(n), \qquad S_n-\hat{S}'_n= O \left(\frac{n^{\beta}}{n^{\beta}+n^{\alpha}} n \right)=O(n^{1+\beta-\alpha})
$$
and
$$
\sigma_n=\Theta(n^{1/2}), \qquad \hat{\sigma}'_n=\Theta \left(\frac{n^{1/2}}{(n^{\alpha}+n^{\beta})^{1/2}} n^{\alpha/2} \right)=\Theta(n^{1/2}).
$$
We then observe that
\begin{align*}
S_n/\sigma_n-\hat{S}'_n/\hat{\sigma}'_n &=S_n/\sigma_n-\hat{S}'_n/\sigma_n+\hat{S}'_n/\sigma_n-\hat{S}'_n/\hat{\sigma}'_n\\
&=(S_n-\hat{S}'_n)/\sigma_n+ \hat{S}'_n(1/\sigma_n-1/\hat{\sigma}'_n)\\
&=(S_n-\hat{S}'_n)/\sigma_n+ \hat{S}'_n \frac{\hat{\sigma}'_n-\sigma_n}{\sigma_n \hat{\sigma}'_n}\\
&=(S_n-\hat{S}'_n)/\sigma_n+ \hat{S}'_n \frac{(\hat{\sigma}'_n)^2-\sigma_n^2}{\sigma_n \hat{\sigma}'_n (\hat{\sigma}'_n+\sigma_n)}\\
&=(S_n-\hat{S}'_n)/\sigma_n+ \hat{S}'_n \frac{O(n^{1+\beta-\alpha})}{\Theta(n^{1/2}) \Theta(n^{1/2}) \Theta(n^{1/2})} \\
&= (S_n-\hat{S}'_n)/\sigma_n +\hat{S}'_n O(n^{\beta-\alpha-1/2}).\\
\end{align*}

For some $\tau < 0$, let $A_1$ denote the event that
$$
\left| \frac{S_n-\hat{S}'_n}{\sigma_n} \right| \geq n^{\tau},
$$
and let $A_2$ denote the event that
$$
\left| \hat{S}'_n \frac{(\hat{\sigma}'_n)^2-\sigma_n^2}{\hat{\sigma}'_n \sigma_n (\hat{\sigma}'_n+\sigma_n)} \right| \geq n^{\tau}.
$$
Then, by the Markov inequality, we have, for any $l \in \mathbb{N}$
$$
P(A_1) = P(|S_n-\hat{S}'_n| \geq n^{\tau+1/2}) \leq \frac{E[|S_n-\hat{S}'_n|^{2l}]}{n^{(\tau+1/2) 2 l}}.
$$
Note that there exist $0 < \theta_1, \theta_2< 1$ such that
\begin{align*}
E[|S_n-\hat{S}'_n|^{2l}] &= E[|S_n-S'_n+O(\theta_1^{q(n)/2}))|^{2l}] \\
&=E[|\eta_1+\eta_2+\cdots+\eta_k|^{2l}]+O(\theta_2^{q(n)/2})\\
&=\sum_{l_1+l_2+\cdots+l_k=2l}O(E[|\eta_1|^{l_1}] E[|\eta_2|^{l_2}] \cdots E[|\eta_k|^{l_k}])+O(\theta_2^{q(n)/2}).
\end{align*}
Then, by Lemmas~\ref{growth-of-even-moments} and~\ref{growth-of-odd-moments}, we obtain, through some further computations, that
$$
E[|S_n-\hat{S}'_n|^{2l}]=O((k n^{\beta})^l).
$$
Now, applying Lemma~\ref{growth-of-even-moments}, ~\ref{growth-of-odd-moments} and Conditions (a) and (c), one can verify that
\begin{equation} \label{A1}
P(A_1) = \frac{O((k n^{\beta})^l)}{n^{(2\tau+1)l}}=O(n^{l(\beta-\alpha-2\tau)}).
\end{equation}
Again, by the Markov inequality, we have, for any $l$,
$$
P(A_2) = P(|\hat{S}'_n| \geq n^{\tau+1/2-\beta+\alpha}) \leq \frac{E[\hat{S}'_n|^{2l}]}{n^{(\tau+1/2-\beta+\alpha) 2 l}}.
$$
Similarly, applying Lemma~\ref{growth-of-even-moments}, ~\ref{growth-of-odd-moments} and Conditions (a), (c), one can verify that
\begin{equation} \label{A2}
P(A_2) = O\left( \frac{(k n^{\alpha})^l}{n^{(2\tau+1-2\beta+2\alpha) l}} \right) = n^{l(-2\tau+2\beta-2\alpha)}.
\end{equation}
Apparently,
$$
P(S_n/\sigma_n \leq x)= P(S_n/\sigma_n \leq x, A_1^c \cap A_2^c)+ P(S_n/\sigma_n \leq x, A_1 \cup A_2),
$$
and
\begin{align*}
P(S_n/\sigma_n \leq x) &= P(\hat{S}'_n/\hat{\sigma}'_n \leq x + \hat{S}'_n/\hat{\sigma}'_n-S_n/\sigma_n )\\
&= P\left(\hat{S}'_n/\hat{\sigma}'_n \leq x + \frac{S_n-\hat{S}'_n}{\sigma_n} + \hat{S}'_n \frac{(\hat{\sigma}'_n)^2-\sigma_n^2}{\hat{\sigma}'_n \sigma_n (\hat{\sigma}'_n+\sigma_n)}\right).
\end{align*}
It then follows from (\ref{A1}) and (\ref{A2}) that for any $x > -\alpha/2$, there exists $\beta > 0$ sufficiently small and $l \in \mathbb{N}$ sufficiently large such that
\begin{equation} \label{A1UA2}
P(S_n/\sigma_n \leq x, A_1 \cup A_2) \leq P(A_1) + P(A_2) = O(n^{-1/4}),
\end{equation}
and
\begin{align*}
P(S_n/\sigma_n \leq x, A_1^c \cap A_2^c) &\geq P(\hat{S}'_n/\hat{\sigma}'_n \leq x-C_1 n^{\tau}, A_1^c \cap A_2^c)\\
&\geq  P(\hat{S}'_n/\hat{\sigma}'_n \leq x-C_1 n^{\tau})+P(A_1^c \cap A_2^c)-1\\
&=P(\hat{S}'_n/\hat{\sigma}'_n \leq x-C_1 n^{\tau})- C_2 n^{-1/4}.
\end{align*}
for some $C_1, C_2 > 0$. On the other hand, it is easy to check that there exists $C_3 > 0$ such that
$$
P(S_n/\sigma_n \leq x, A_1^c \cap A_2^c) \leq P(\hat{S}'_n/\hat{\sigma}'_n \leq x+ C_3 n^{\tau}).
$$
Noticing that
$$
|P(S_n/\sigma_n \leq x) - G(x)| \leq \max \{ P(\hat{S}'_n/\hat{\sigma}'_n \leq x+C_3 n^{x})-G(x), G(x)- P(\hat{S}'_n/\hat{\sigma}'_n \leq x-C_1 n^{x})+C_2 n^{-1/4}\},
$$
and applying Lemma~\ref{SnPrime}, we derive
$$
|P(\hat{S}'_n/\hat{\sigma}'_n \leq x+C_3 n^{\tau})-G(x)| \leq |P(\hat{S}'_n/\hat{\sigma}'_n \leq x+C_3 n^{\tau})-G(x+C_3 n^{\tau})|
$$
$$
+|G(x+C_3 n^{\tau})-G(x)|=O(n^{-1/2+\alpha/2})+O(n^{\tau}),
$$
and similarly,
$$
|G(x)- P(\hat{S}'_n/\hat{\sigma}'_n \leq x-C_1 n^{\tau})+C_2 n^{-1/4}| = O(n^{-1/2+\alpha/2})+O(n^{\tau})+O(n^{-1/4}).
$$
Setting $\alpha=1/2$, $\tau$ slightly larger than $-1/4$, and choosing $\beta > 0$ sufficiently small, we then have established the theorem.
\end{proof}

\begin{rem}
If Condition (II) fails, i.e., $\lim_{n \to \infty} \sigma_n^2/n=0$, then a CLT of degenerated form holds for $(X_i, i \in \mathbb{N})$; more precisely, the distribution of $(X_1+X_2+\cdots+X_n)/\sqrt{n}$ converges to that of a centered normal distribution with variance $0$, i.e., a point mass at $0$, as $n \rightarrow \infty$. This is can be readily checked since for any $\eps > 0$, by the Markov inequality, we have
$$
P(|(X_1+X_2+\cdots+X_n)|/\sqrt{n} \geq \eps|) \leq \sigma_n^2/(n\eps^2) \rightarrow 0 \mbox{ as } n \rightarrow \infty.
$$
\end{rem}

\subsection{Proof of Theorem~\ref{ASIP-Theorem}}

Consider the following Bernstein blocking method with variable block lengths: we consecutively partition the partial sum $S_n$ into blocks $\eta_1, \zeta_1, \eta_2, \zeta_2, \ldots$ such that $\eta_j$ is of length $q_j = q_j(n) \triangleq j^{\beta}$ and $\zeta_j$ is of length $p_j =p_j(n) \triangleq j^{\alpha}$. Similarly as in the proof of Theorem~\ref{CLT-Theorem}, we have
$$
S_n=S^*_n+S'_n,
$$
where $S^*_n$ is the sum of all feasible $\eta$-blocks and $S'_n$ is the sum of all feasible $\zeta$-blocks.
Let $\mathcal{L}_i$ denote the $\sigma$-algebra generated by all $X_j$'s contained in $\zeta_i$. It is well known that $S'_n$ can be approximated using a Martinagle in the following manner
$$
\zeta_i=\xi_i+\nu_i-\nu_{i+1},
$$
where
$$
\xi_i=\sum_{k=0}^{\infty} (E[\zeta_{i+k}|\mathcal{L}_{i}]-E[\zeta_{i+k}|\mathcal{L}_{i-1}])
$$
is a Martingale difference sequence, and
$$
\nu_i=\sum_{k=0}^{\infty} E[\zeta_{i+k}|\mathcal{L}_{i-1}].
$$

Similarly as in the proof of Theorem~\ref{CLT-Theorem}, we truncate $\zeta$-blocks in the following way: Consider a $\zeta$-block taking the following form
$$
\zeta_i=X_{j_1}+X_{j_1+1}+\cdots+X_{j_2}
$$
For any $j= j_1, j_1+1, \cdots, j_2$, define
$$
\hat{X}_{j}=f(Z_j|Z_{j_1-q_j/2}^{j-1})-E[f(Z_j|Z_{j_1-q_j/2}^{j-1})],
$$
and further
$$
\hat{\zeta}_i=\hat{X}_{j_1}+\hat{X}_{j_1+1}+\cdots+\hat{X}_{j_2}.
$$

Before proving Theorem~\ref{ASIP-Theorem}, we need to establish several lemmas. The following lemma states that $\nu_i$ is sub-exponentially small with respect to $i$.

\begin{lem} \label{sub-exponentially-small}
There exist $C > 0$, $0 < \theta < 1$ and $0 < \delta < \beta$ such that for all $i$,
$$
|\nu_i| \leq C \theta^{i^{\delta}}
$$
\end{lem}

\begin{proof}
Recall that for some $0 < \rho < 1$,
$$
\zeta_i-\hat{\zeta}_i=O(\rho^{q_i/2}).
$$
We then have
\begin{align*}
\nu_i=\sum_{k=0}^{\infty} E[\zeta_{i+k}|\mathcal{L}_{i-1}] &=\sum_{k=0}^{\infty} (E[\hat{\zeta}_{i+k}|\mathcal{L}_{i-1}]+O(p_{i+k} \rho^{q_{i+k}/2})) \\
&=\sum_{k=0}^{\infty}(E[\hat{\zeta}_{i+k}]+O(\lambda^{q_{i+k}/2}) E[|\hat{\zeta}_{i+k}|]) +\sum_{k=0}^{\infty} O(p_{i+k} \rho^{q_{i+k}/2}).
\end{align*}
Noting that $E[\hat{\xi}_{i+k}]=0$ and the constants in the above O-terms are independent of $k$, we conclude that $\nu_i$ is sub-exponentially small with respect to $i$.
\end{proof}

By the classical Skorokhod representation theorem (see~\cite{bi95}), there exist non-negative random variables $T_i$ such that for all feasible $M$,
$$
\sum_{i \leq M} \xi_i=B(\sum_{i \leq M} T_i) \mbox{  a.s. }
$$
and
$$
E[T_i|\mathcal{L}_{i-1}] = E[\xi_i^2|\mathcal{L}_{i-1}] \mbox{ a.s.}, \quad E[T_i^p] = O(E[|\xi_i|^{2p}]) \mbox{ for each $p > 1$. }
$$
Let $M_N$ denote the index of the $\zeta$-block or the $\eta$-block containing $X_N$. Then, depending on $X_N$ is contained in a $\zeta$-block or a $\eta$-block, we have either
$$
\sum_{i=1}^{M_N-1} i^{\alpha}+\sum_{i=1}^{M_N} i^{\beta} \leq N \leq \sum_{i=1}^{M_N} i^{\alpha}+\sum_{i=1}^{M_N} i^{\beta},
$$
or
$$
\sum_{i=1}^{M_N-1} i^{\alpha}+\sum_{i=1}^{M_N-1} i^{\beta} \leq N \leq \sum_{i=1}^{M_N-1} i^{\alpha}+\sum_{i=1}^{M_N} i^{\beta}.
$$
Using the fact that
$$
\int_0^n x^{\alpha} dx \leq 1^{\alpha}+2^{\alpha}+\cdots+n^{\alpha} \leq \int_1^{n+1} x^{\alpha} dx,
$$
we deduce that
$$
\frac{n^{\alpha+1}}{\alpha+1} \leq 1^{\alpha}+2^{\alpha}+\cdots+n^{\alpha} \leq \frac{(n+1)^{\alpha+1}-1}{\alpha+1}.
$$
We then have either
$$
\frac{(M_N-1)^{\alpha+1}}{\alpha+1}+\frac{M_N^{\beta+1}}{\beta+1} \leq N \leq \frac{(M_N+1)^{\alpha+1}-1}{\alpha+1}+\frac{(M_N+1)^{\beta+1}-1}{\beta+1},
$$
or
$$
\frac{(M_N-1)^{\alpha+1}}{\alpha+1}+\frac{(M_N-1)^{\beta+1}}{\beta+1} \leq N \leq \frac{M_N^{\alpha+1}-1}{\alpha+1}+\frac{(M_N+1)^{\beta+1}-1}{\beta+1}.
$$
Apparently, we have, for either of the above cases,
$$
M_N = \Theta(N^{1/(\alpha+1)}).
$$
As elaborated in~\cite{ph75}, a somewhat standard procedure can be followed to establish an almost sure invariance principle. For Theorem~\ref{ASIP-Theorem} in this paper, it suffices to prove that
\begin{enumerate}
\item for any $\eps > 0$,
\begin{equation} \label{smalls-added-up}
\sum_{i=1}^{M_N} \eta_i = O(N^{1/3+\eps}) \mbox{ a.s.};
\end{equation}
\item for any $\eps > 0$,
\begin{equation} \label{T-N}
\sum_{j=1}^{M_N} T_j=\sigma^2 N+O(N^{2/3+\eps}) \mbox{ a.s.},
\end{equation}
as $N$ tends to infinity.
\end{enumerate}
We will establish (\ref{smalls-added-up}) in Lemma~\ref{etas-added-up}. To establish (\ref{T-N}), consider the following decomposition
$$
\sum_{i=1}^{M_N} T_i - \sigma^2 N = \sum_{i=1}^{M_N} (T_i-E[T_i|\mathcal{L}_{i-1}])+\sum_{i=1}^{M_N} (E[\xi_i^2|\mathcal{L}_{i-1}]-\xi_i^2)+(\sum_{i=1}^{M_N} \xi_i^2- \sigma^2 N).
$$
It is then clear that we only need to prove all the above three terms are of $O(N^{2/3+\eps})$, for any $\eps > 0$.

We need the following well-known lemma, whose proof can be found in~\cite{ph75}.

\begin{lem} \label{Theorem-A1}
Let $\{x_j\}$ be a sequence of centered random variables with finite second moments. Suppose that there exists a constant $s > 0$ such that all integers $k \geq j$,
$$
E[(\sum_{i=j}^{k} x_i)^2] = O(k^{s}-j^{s}).
$$
Then for each $\delta > 0$, we have
$$
\sum_{i=1}^N x_j = O(N^{s/2 } \log^{2+\delta} N)  \mbox{ a.s. }
$$
\end{lem}

The following lemma establishes (\ref{smalls-added-up}).
\begin{lem} \label{etas-added-up}
With probability $1$,
$$
\sum_{i=1}^{M_N} \eta_i = O(N^{1/3+\eps}),
$$
for any $\eps > 0$.
\end{lem}

\begin{proof}
Note that for any $j, k$,
$$
E[(\sum_{i=j}^{k} \eta_i)^2]=\sum_{i=j}^k E[\eta_i^2]+ 2 \sum_{i < j} E[\eta_i \eta_j].
$$
First, notice that an argument parallel to the proof for Part $2$ of Lemma~\ref{two} with Conditions (a) and (c) implies that $E[\eta_i \eta_j]$ sub-exponentially small in $j-i$, and thus
$$
\sum_{i < j} E[\eta_i \eta_j]=O(1).
$$
Applying Lemma~\ref{VarianceLemma}, we have for some small $\eps_0 > 0$,
$$
E[(\sum_{i=j}^{k} \eta_i)^2]=\sum_{i=j}^k (\sigma^2 i^{\beta}+O(i^{\eps_0})) + O(1)=O(k^{\beta+1}-j^{\beta+1})+O(k^{\eps_0+1}-j^{\eps_0+1}).
$$
It then follows from Lemma~\ref{Theorem-A1} that for any $\beta' > \beta$, $\eps'_0 > \eps_0$,
$$
\sum_{i=1}^{M_N} \eta_i=O(M_N^{(\beta'+1)/2})+O(M_N^{(\eps'_0+1)/2})=O(N^{(\beta'+1)/(2(\alpha+1))})+O(M_N^{(\eps'_0+1)/(2(\alpha+1))}),
$$
where we have applied the fact that $M_N=\Theta(N^{1/(\alpha+1)})$. Choosing $\beta, \eps_0, \beta', \eps'_0 > 0$ sufficiently small and setting $\alpha=1/2$, the lemma then immediately follows.
\end{proof}

The following three lemmas collectively establish (\ref{T-N}).
\begin{lem} \label{the-first-term} 
With probability $1$,
$$
\sum_{i=1}^{M_N} \xi_i^2-\sigma^2 N=O(N^{2/3+\eps})
$$
\end{lem}
for any $\eps > 0$.

\begin{proof}
Note that by Lemma~\ref{sub-exponentially-small}, $\zeta_i$ and $\xi_i$ are sub-exponentially close. So, we only need to prove that
$$
\sum_{i=1}^{M_N} \zeta_i^2-\sigma^2 N=O(N^{2/3+\eps}) \mbox{ a.s. }
$$
for any $\eps > 0$.

Depending on whether $X_N$ is contained in a $\zeta$-block or a $\eta$-block, we have either
$$
-M_N^{\alpha}+\sum_{i=1}^{M_N} i^{\beta} \leq N- \sum_{i=1}^{M_N} i^{\alpha} \leq \sum_{i=1}^{M_N} i^{\beta},
$$
which implies that
$$
-M_N^{\alpha}+\frac{M_N^{\beta+1}}{\beta+1} \leq N-\sum_{i=1}^{M_N} i^{\alpha} \leq \frac{(M_N+1)^{\beta+1}-1}{\beta+1},
$$
or
$$
\sum_{i=1}^{M_N-1} i^{\beta}-M_N^{\alpha} \leq N-\sum_{i=1}^{M_N} i^{\alpha}
\leq \sum_{i=1}^{M_N} i^{\beta}-M_N^{\alpha},
$$
which implies that
$$
\frac{(M_N-1)^{\beta+1}}{\beta+1} - M_N^{\alpha} \leq N-\sum_{i=1}^{M_N} i^{\alpha} \leq \frac{(M_N+1)^{\beta+1}-1}{\beta+1} - M_N^{\alpha}.
$$
In any case, applying Lemma~\ref{VarianceLemma}, we have for some small $\eps_0 > 0$,
\begin{align*}
E[\sum_{i=1}^{M_N} \zeta_i^2]-\sigma^2 N &=\sum_{i=1}^{M_N} (\sigma^2 i^{\alpha}+O(i^{\eps_0}))-\sigma^2 N \\
& = \sigma^2 (\sum_{i=1}^{M_N} i^{\alpha}-N)+O(\sum_{i=1}^{M_N} i^{\eps_0})\\
& = O(M_N^{\alpha})+O(M_N^{\beta+1})+O(M_N^{\eps_0+1})\\
& = O(N^{\alpha/(\alpha+1)})+O(N^{(\beta+1)/(\alpha+1)})+O(N^{(\eps_0+1)/(\alpha+1)}),
\end{align*}
where we have applied the fact that $M_N=\Theta(N^{1/(\alpha+1)})$. Choosing $\beta, \eps_0 > 0$ small enough and setting $\alpha=1/2$, we then have
$$
E[\sum_{i=1}^{M_N} \zeta_i^2]-\sigma^2 N=O(N^{2/3+\eps})
$$
for any $\eps > 0$.

So, to prove the lemma, it suffices to prove that with probability $1$,
$$
\sum_{i=1}^{M_N} (\zeta_i^2-E[\zeta_i^2])=O(N^{2/3+\eps})
$$
for any $\eps > 0$. Using Conditions (a) and (c), we derive that with probability $1$,
\begin{align*}
|E[\zeta_i^2]-E[\zeta_i^2|\mathcal{L}_{i-1}]|
&= |E[\hat{\zeta}_i^2]-E[\hat{\zeta}_i^2|\mathcal{L}_{i-1}]|+O(E[|\hat{\zeta}_i|]i^{\alpha} \rho^{q_i/2})+O(i^{2 \alpha} \rho^{q_i}) \\
&=O(E[\hat{\zeta}_i^2] \lambda^{q_{i-1}})+O(E[|\hat{\zeta}_i|]i^{\alpha} \rho^{q_i/2})+O(i^{2 \alpha} \rho^{q_i}).
\end{align*}
Applying Lemma~\ref{growth-of-even-moments}, we then have for any $j, k$,
\begin{align*}
E[(\sum_{i=j}^{k} (\zeta_i^2-E[\zeta_i^2]))^2]
& = E[(\sum_{i=j}^k (\zeta_i^2-E[\zeta_i^2|\mathcal{L}_{i-1}]))^2]+O(1)\\
& = \sum_{i=j}^k E[(\zeta_i^2-E[\zeta_i^2|\mathcal{L}_{i-1}])^2]+O(1)\\
& \leq \sum_{i=j}^k E[\zeta_i^4]+O(1) \\
& = O(j^{2 \alpha}+(j+1)^{2 \alpha}+\cdots+k^{2 \alpha})+O(1) \\
& = O(k^{2 \alpha+1}-j^{2 \alpha+1}),
\end{align*}
where we have used the fact that for $i_1 \neq i_2$,
$$
E[(\zeta_{i_1}^2-E[\zeta_{i_1}^2|\mathcal{L}_{i_1-1}]) (\zeta_{i_2}^2-E[\zeta_{i_2}^2|\mathcal{L}_{i_2-1}])]=0.
$$
Applying Lemma~\ref{Theorem-A1}, we then have, for any $\alpha' > \alpha$,
$$
\sum_{i=1}^{M_N} (\zeta_i^2-E[\zeta_i^2]) = O(M_N^{(2\alpha'+1)/2})=O(N^{(2\alpha'+1)/(2(1+\alpha))}) \mbox{ a.s. }
$$
Setting $\alpha=1/2$ and choosing $\alpha'$ slightly larger than $1/2$, we then have proven the lemma.
\end{proof}

\begin{lem} \label{the-second-term} 
With probability $1$,
$$
\sum_{i=1}^{M_N} (E[\xi_i^2|\mathcal{L}_{i-1}]-\xi_i^2)=O(N^{2/3+\eps})
$$
\end{lem}
for any $\eps > 0$.

\begin{proof}
Note that by Lemma~\ref{sub-exponentially-small}, $\zeta_i$ and $\xi_i$ are sub-exponentially close. So, we only need to prove that
$$
\sum_{i=1}^{M_N} (E[\zeta_i^2|\mathcal{L}_{i-1}]-\zeta_i^2)=O(N^{2/3+\eps}) \mbox{ a.s. }
$$
for any $\eps > 0$. But this has been established in the proof of the previous lemma.
\end{proof}

\begin{lem} With probability $1$,
$$
\sum_{i=1}^{M_N} (T_i-E[T_i|\mathcal{L}_{i-1}])=O(N^{2/3+\eps})
$$
for any $\eps > 0$.
\end{lem}

\begin{proof}
Similarly as in the proof of Lemma~\ref{the-first-term}, we have that for any $j, k$,
\begin{align*}
E[(\sum_{i=j}^{k} (T_i-E[T_i|\mathcal{L}_{i-1}]))^2] & =\sum_{i=j}^{k} E[(T_i-E[T_i|\mathcal{L}_{i-1}])^2]\\
& \leq \sum_{i=j}^{k} E[T_i^2]\\
& \leq \sum_{i=j}^{k} E[\zeta_i^4] \\
& \leq O(j^{2\alpha}+(j+1)^{2\alpha}+\cdots+k^{2\alpha})\\
& =O(k^{2\alpha+1}-j^{2\alpha+1}).
\end{align*}
Then, similarly as in the proof of Lemma~\ref{the-second-term}, we deduce that
$$
\sum_{i=1}^{M_N} (T_i-E[T_i|\mathcal{L}_{i-1}]) = O(N^{2/3+\eps}) \mbox{ a.s. }
$$
for any $\eps > 0$.
\end{proof}

\subsection{Proof of Theorem~\ref{Chernoff-Bound-Theorem}}
In this proof, we assume the Bernstein blocking as in Theorem~\ref{CLT-Theorem}. Notice that
$$
P(S_n/n \geq \eps)=P(S_n \geq n \eps)=P(\hat{S}'_n+S_n-\hat{S}'_n \geq n \eps)=P(\hat{S}'_n \geq n \eps - (S_n-\hat{S}'_n)).
$$
Notice that $S_n-\hat{S}'_n = O(n^{1+\alpha-\beta})$, so we have
\begin{equation} \label{MI-Snn}
P(S_n/n \geq \eps) \leq P(\hat{S}'_n \geq n \eps') = P(t \hat{S}'_n/p \geq t n \eps'/p) \leq \frac{E[e^{t \hat{S}'_n/p]}}{e^{t n \eps'/p }},
\end{equation}
for some $0 < \eps' < \eps$. Applying Condition (a), we then have
\begin{equation} \label{first-iteration}
E[e^{t \hat{S}'_n/p}]=E[e^{t \sum_{i=1}^{k-1} \hat{\zeta}_i/p} e^{t \hat{\zeta}_k/p}] = (1 + O(\lambda^{q(n)/2})) E[e^{t \sum_{i=1}^{k-1} \zeta_i/p}] E[e^{t \hat{\zeta}_k}].
\end{equation}
An iterative application of (\ref{first-iteration}) gives us that for any $0 < t < 1$
\begin{eqnarray} \label{Sn-prime}
\nonumber E[e^{t \hat{S}'_n/p}]&=&E[e^{t \sum_{i=1}^k \hat{\zeta}_i/p}] \\
&=& (1 + O(\lambda^{q(n)/2}))^{k-1} (E[e^{t \hat{\zeta}_1/p}])^k,
\end{eqnarray}
as $n$ goes to infinity. If Condition (d) holds, by Lemma~\ref{VarianceLemma}, as $n$ goes to infinity (and hence $p, q$ go to infinity), we have
$$
E[\hat{\zeta}_1^2]/p^2= o(1),
$$
which trivially holds when Condition (d) fails. It then follows that for any $0 < t < 1$,
$$
E[e^{t \hat{\zeta}_1/p}]=1+ o(1)t^2,
$$
and furthermore, for $t > 0$ sufficiently small, we have
\begin{equation} \label{less-than-one}
\frac{E[e^{t \hat{\zeta}_1/p}]}{e^{t \eps'}}=\frac{1+o(1)t^2}{1+t\eps' + O(1) t^2} < 1.
\end{equation}
Now, from (\ref{MI-Snn}), (\ref{Sn-prime}) and (\ref{less-than-one}), we deduce that for any $\eps > 0$, there exists $\eps' > 0$ such that
\begin{eqnarray}
\nonumber P(S_n/n \geq \eps) & \leq & \frac{E[e^{t \hat{S}'_n/p]}}{e^{t n \eps'/p }}\\
\nonumber &\leq & (1+O(\lambda^{q(n)/2}))^k (E[e^{t \zeta_1/p}]/e^{t \eps'})^{k}.
\end{eqnarray}
Notice that for sufficiently large $n$, we have
$$
(1+O(\lambda^{q(n)/2})) E[e^{t \zeta_1/p}]/e^{t \eps'} < 1,
$$
which, together with $\alpha > 0$ chosen sufficiently small, we conclude that for any $x,\eps > 0$, there exists $0 < \gamma < 1$ such that
$$
P(S_n/n \geq x) = O(\gamma^{n^{1-\eps}}).
$$
The proof is then complete.

\subsection{Alternatives for Condition (d)}

Note that for the case $Z$ is in fact a Markov chain, a rather explicit alternative condition for Condition (d) has been derived in~\cite{yu53}. This section only assumes Conditions (a), (b), (c) and gives alternatives for Condition (d) provided Conditions (a), (b), (c) are satisfied.

Let $(\Omega, \mathcal{F}, P)$ be the probability space on which $Z$ is defined, and let $H_0=H(Z_k, k \in \mathbb{Z})$ be the subspace of $\mathcal{L}^2(\mathcal{F})$ spanned by the equivalence classes of the random variables $Z_k$, $k \in \mathbb{Z}$, with inner product defined as
$$
<V, W>=E[V W],
$$
for any $V, W \in H_0$.

\begin{lem} \label{dichotomy}
If $\liminf_{n \rightarrow \infty} E[S_n^2] < \infty$, then there exists a sequence of random variables $(V_i, i \in \mathbb{N})$ such that $X_i=V_i-V_{i+1}$ with $E[V_i^2]=O(1)$ uniformly for all $i$, and thus $\sup_n E[S_n^2] < \infty$.
\end{lem}

\begin{proof}

Let $Q$ be an infinite subset of $\mathbb{N}$ such that $\sup_{n \in Q}E[S_n^2] < \infty$. Applying Condition (c), we have for any $n, m$,
$$
E[(S_{n+m}-S_{m})^2]=E[(\sum_{i=m+1}^{n+m} g(Z_1^i))^2]=E[(\sum_{i=m+1}^{n+m} (g(Z_{m+1}^i)+O(\rho^{i-m-1})))^2]
$$
$$
=E[S_n^2]+O(E[|S_n|])+O(1) = E[S_n^2]+O(E[S_n^2]^{1/2})+O(1).
$$
We then deduce that there exists $C > 0$ such that for all $i \in \mathbb{N}$,
$$
\sup_{n \in Q} E[(S_{n+i-1}-S_{i-1})^2] \leq C,
$$
where $S_0$ is interpreted as $0$. It follows from the Banach-Alaoglu theorem (which states that every bounded and closed set in a Hilbert space is weakly compact; see Section $3.15$ of~\cite{ru91}) that for any $i \in \mathbb{N}$, there exists $V_i \in H_0$ with $E[V_i^2] \leq C$, and $Q_i$, an infinite subset of $Q$ such that for all $W \in H_0$,
$$
\lim_{n \to \infty, n \in Q_i} <W, S_{n+i-1}-S_{i-1}>=<W, V_i>;
$$
here, without loss of generality, we can assume that $Q_{i+1} \subset Q_i$ for all $i$. Then one verifies that for any $W \in H_0$, we have that for any $i$,
$$
\hspace{-1cm} <W, X_i-V_i+V_{i+1}>=\lim_{n \rightarrow \infty, n \in Q_{i+1}} <W, X_i-(S_{n+i-1}-S_{i-1})+(S_{n+i}-S_i)>=\lim_{n \rightarrow \infty, n \in Q} <W, X_{n+i}>=0,
$$
where we have applied Lemma~\ref{two} for the last equality. Choosing $W=X_i-V_i+V_{i+1}$, we then obtain that
$$
\|X_i-V_i+V_{i+1}\|_2=0,
$$
which implies that
$$
X_i=V_i-V_{i+1}, \mbox{ a.s. }
$$
It then follows that
$$
E[S_n^2]=E[(V_1-V_{n+1})^2]=E[V_1^2]+E[V_{n+1}^2]-2E[V_1V_{n+1}],
$$
which, together with $E[V_i^2] \leq C$, implies the theorem.

\end{proof}

A sequence of positive numbers, $(h(i), i \in \mathbb{N})$, is said to be {\em slowly varying} if for every positive integer $m$,
$$
\lim_{n \to \infty} h(mn)/h(n)=1,
$$
and it is said to be {\em slowly varying in the strong sense} if
$$
\lim_{m \to \infty} \frac{\min_{m \leq n \leq 2m} h(n)}{\max_{m \leq n \leq 2m} h(n)}=1.
$$

\begin{lem} \label{not-strong}
If $\lim_{n \to \infty} E[S_n^2]=\infty$, then $E[S_n^2]=n h(n)$, where $(h(i), i \in \mathbb{N})$ is a sequence of slowly varying positive numbers.
\end{lem}

\begin{proof}
We only need to show that for every positive integer $l$,
$$
\lim_{n \rightarrow \infty} \sigma_{ln}^2/\sigma_n^2=l.
$$
Following~\cite{li96}, we use the Bernstein blocking method in the following way: We consecutively partition the partial sum $S_{ln}$ into blocks $\zeta_1, \eta_1, \zeta_2, \eta_2, \ldots$ such that each $\zeta_i$ is of length $n$ and each $\eta_i$ is of length $r=\lfloor \log \sigma_n^2 \rfloor$. In other words, for any feasible $i$,
$$
\zeta_i=\sum_{s=1}^n X_{(i-1)n+(i-1)r+s}, \eta_i=\sum_{s=1}^r X_{in+(i-1)r+s}.
$$
Now,
$$
\sigma_{ln}^2=E[S_{ln}^2]=\sum_{j=1}^l E[\zeta_j^2]+2 \sum_{i \neq j} E[\zeta_i \zeta_j]+\sum_{i, j} E[\zeta_i \eta_j]+\sum_{i, j} E[\eta_i \eta_j].
$$
It follows from Lemma~\ref{VarianceLemma} that for any $j$,
\begin{equation} \label{thesame}
E[\zeta_j^2]=\sigma_n^2+O(\sigma_n).
\end{equation}
Using an argument similar to the proof for Part $2$ of Lemma~\ref{two}, one has that there exists $0 < \theta < 1$ such that for $i \neq j$,
$$
|E[\zeta_i \zeta_j]| = O(\theta^{\lfloor \log \sigma_n^2 \rfloor} \sigma_n^2),
$$
where we also used (\ref{thesame}). Using the Schwartz inequality and (\ref{thesame}), we also have
$$
|E[\zeta_i \eta_j]| \leq E[\zeta_i^2]^{1/2} E[\eta_j^2]^{1/2} =O(\sigma_n \sigma_r)=O(\sigma_n \log \sigma_n),
$$
and
$$
|E[\eta_i \eta_j]| \leq O(\sigma_r^2)=O((\log \sigma_n)^2).
$$
It then follows that for any positive integer $l$,
$$
\sigma_{ln}^2=l\sigma_n^2+o(\sigma_n^2),
$$
which immediately implies the lemma.
\end{proof}

\begin{lem} \label{strong}
If $\lim_{n \to \infty} E[S_n^2]=\infty$, then $E[S_n^2]=n h(n)$, where $(h(i), i \in \mathbb{N})$ is a sequence of slowly varying positive numbers in the strong sense.
\end{lem}

\begin{proof}
Note that by Lemma~\ref{VarianceLemma}, we have that for any $j$,
\begin{equation} \label{ratio-is-1}
\lim_{n \to \infty} \frac{E[(S_{n+j}-S_j)^2]}{E[S_n^2]}=1,
\end{equation}
uniformly in $j$. The lemma then follows from (\ref{ratio-is-1}), Lemma~\ref{not-strong} and an almost the same proof for Theorem $8.13$ of~\cite{br07}.
\end{proof}

The following lemma is well-known; see, e.g., Proposition $0.16$ in~\cite{br07}.
\begin{lem}  \label{slow-and-slower}
Suppose $(h(n), n \in \mathbb{N})$ is a sequence of positive numbers which is slowly varying in the strong sense. Then for every $\eps > 0$, one has that $n^{\eps} h(n) \rightarrow \infty$ as $n \rightarrow \infty$.
\end{lem}

\begin{lem} \label{positive-sigma}
If $\lim_{n \rightarrow \infty} E[S_n^2] = \infty$, then $\sigma > 0$.
\end{lem}

\begin{proof}
Assume, by contradiction, that $\sigma=0$. Since $\lim_{n \rightarrow \infty} E[S_n^2] = \infty$, we deduce, by Lemma~\ref{strong}, that $E[S_n^2]/n$ is slowly varying in the strong sense. Then, by Lemma~\ref{slow-and-slower}, for any $\alpha > 0$, $n^{\alpha} E[S_n^2]/n \rightarrow \infty$ as $n \rightarrow \infty$. However, by Lemma~\ref{VarianceLemma}, when $\sigma=0$, $n^{\alpha} E[S_n^2]/n \rightarrow 0$ as $n \rightarrow \infty$ for any $0 < \alpha < 1$, which is a contradiction.
\end{proof}

The following theorem immediately follows from Lemma~\ref{dichotomy} and Lemma~\ref{positive-sigma}, which gives alternatives for Condition (d) given Conditions (a), (b) and (c) are satsified.
\begin{thm}
Under Conditions (a), (b) and (c), the following statements are equivalent
\begin{enumerate}
\item $\sigma > 0$.
\item $\lim_{n \rightarrow \infty} E[S_n^2] = \infty$.
\item $\limsup_{n \rightarrow \infty} E[S_n^2] = \infty$.
\end{enumerate}
\end{thm}

\section{Proofs of the Main Results}
Unless specified otherwise, all the lemmas in this section only assume Condition (I).

For each $z \in \mathcal{Z}$, let $\Delta_z$ denote the matrix such that $\Delta_z(i, j)=\Delta(i, j) p(z|j)$ for all feasible $i, j$; obviously $\sum_{z \in \mathcal{Z}} \Delta_z = \Delta$. One also observes that for any $z_{m_1}^{m_2}$,
$$
p(z_{m_1}^{m_2})=\pi \Delta_{z_{m_1}^{m_2}} \mathbf{1},
$$
where $\pi$ is the stationary vector of $Y$, $\mathbf{1}$ denotes the all one column vector and $\Delta_{z_{m_1}^{m_2}} \triangleq \Delta_{z_{m_1}} \Delta_{z_{m_1+1}} \cdots \Delta_{z_{m_2}}$. Since $\Delta$ is irreducible and aperiodic, $\Delta^{m_2}_{m_1}$ is strictly positive if $m_2-m_1$ is large enough. Notice that $\Pi$ is strictly positive, by reblocking the process $Z$ if necessary, we may assume that all $\Delta_z$ are positive. It then follows from the argument in~\cite{gm05} and the quotient rule (for taking the derivatives) that
\begin{lem} \label{uniformly-bounded}
For any $l \geq 0$ and any compact subset $\Omega_0 \subset \Omega$, there exists $C > 0$ such that for any $z_{-n}^0$ and any $\theta \in \Omega_0$,
$$
|D^l_{\theta} \log p(z_0|z_{-n}^{-1})| < C.
$$
\end{lem}

For $\delta > 0$, let $\mathbb{C}_{\mathbb{R}^+}[\delta]$ denote the ``relative'' $\delta$-neighborhood of $\mathbb{R}^+ \triangleq \{x \in \mathbb{R}: x > 0\}$ within $\mathbb{C}$, i.e.,
$$
\mathbb{C}_{\mathbb{R}^+}[\delta]=\{z \in \mathbb{C}: |z-x| \leq \delta x, \mbox{ for some } x > 0\}.
$$
Let $\mathbb{C}^m_{\theta}(r)$ denote the $r$-neighborhood of $\theta$ in $\mathbb{C}^m$.
It turns out that for $r > 0$ small enough, $p^{\theta}(z_0|z_{-n}^{-1})$, $H^{\theta}(Z_0|Z_{-n}^{-1})$ can be analytically continued to $p^{\tilde{\theta}}(z_0|z_{-n}^{-1})$, $H^{\tilde{\theta}}(Z_0|Z_{-n}^{-1})$ for all $\tilde{\theta} \in \mathbb{C}^m_{\theta}(r)$, respectively. With the fact that an $n \times n$ positive matrix induces a contraction mapping on the interior of the $(n-1)$-dimensional real simplex under the Hilbert metric~\cite{se80}, the following lemma has been established in~\cite{gm05} (see also a more direct proof in~\cite{hm09a} using a complex Hilbert metric).
\begin{lem}  \label{Complex-ExpoForgetting}
\begin{enumerate}
\item For any $\delta > 0$, there exists $r > 0$ such that for any $\tilde{\theta} \in \mathbb{C}_{\theta_0}^{m}(r)$ and for any $z_{-n}^0 \in \mathcal{Z}^{n+1}$,
$$
p^{\tilde{\theta}}(z_0|z_{-n}^{-1}) \in \mathbb{C}_{\mathbb{R}^+}[\delta].
$$
\item There exist $C > 0$, $0 < \rho < 1$ and $r > 0$ such that for any two hidden Markov sequences $z_{-m}^0, \hat{z}_{-\hat{m}}^0$ with $z_{-n}^0=\hat{z}_{-n}^0$ (here $m, \hat{m} \geq n \geq 0$) and all $\tilde{\theta} \in \mathbb{C}^m_{\theta}(r)$, we have
$$
|p^{\tilde{\theta}}(z_0|z_{-m}^{-1}) - p^{\tilde{\theta}}(\hat{z}_0|\hat{z}_{-\hat{m}}^{-1}) | \leq C \rho^n,
$$
and
$$
|\log p^{\tilde{\theta}}(z_0|z_{-m}^{-1}) - \log p^{\tilde{\theta}}(z_0|z_{-\hat{m}}^{-1}) | \leq C \rho^n, \quad |E_{\theta_0}[\log p^{\tilde{\theta}}(Z_0|Z_{-m}^{-1})]-E_{\theta_0}[\log p^{\tilde{\theta}}(Z_0|Z_{-\hat{m}}^{-1})]| \leq C \rho^n.
$$
\end{enumerate}
\end{lem}
\noindent Together with the Cauchy integral formula, the above lemma immediately implies the following corollary.
\begin{co} \label{Real-ExpoForgetting}
For any $l \geq 0$ and any compact subset $\Omega_0 \subset \Omega$, there exist $C > 0$, $0 < \rho < 1$ such that for any two hidden Markov sequences $z_{-m}^0, \hat{z}_{-\hat{m}}^0$ with $z_{-n}^0=\hat{z}_{-n}^0$ (here $m, \hat{m} \geq n \geq 0$) and any $\theta \in \Omega_0$,
$$
|D^l_{\theta} p^{\theta}(z_0|z_{-m}^{-1}) - D^l_{\theta} p^{\theta}(\hat{z}_0|\hat{z}_{-\hat{m}}^{-1}) | \leq C \rho^n,
$$
and
$$
\hspace{-1cm} |D^l_{\theta} \log p^{\theta}(z_0|z_{-m}^{-1}) - D^l_{\theta} \log p^{\theta}(z_0|z_{-\hat{m}}^{-1}) | \leq C \rho^n, \quad |E_{\theta_0} [D^l_{\theta} \log p^{\theta}(Z_0|Z_{-m}^{-1})]-E_{\theta_0}[D^l_{\theta} \log p^{\theta}(Z_0|Z_{-\hat{m}}^{-1})]| \leq C \rho^n.
$$
\end{co}

It is well known~\cite{br05} that a finite-state irreducible and aperiodic Markov chain is a $\psi$-mixing sequence, and the corresponding $\psi(n)$ exponentially decays as $n \to \infty$. The following lemma asserts that under Condition (I), $Z$ is a $\psi$-mixing sequence and the corresponding $\psi(n)$ exponentially decays as $n \to \infty$. An excellent survey on various mixing sequences can be found in~\cite{br05}; for a comprehensive exposition to the vast literature on this subject, we refer to~\cite{br07}.
\begin{lem} \label{Mixing}
$Z^{\theta}$ is a $\psi$-mixing sequence, and for any compact subset $\Omega_0 \subset \Omega$, there exist $C > 0$ and $0 < \lambda < 1$ such that for any positive $n$ and any $\theta \in \Omega_0$,
$$
\psi(n) \leq C \lambda^n.
$$
\end{lem}

\begin{proof}

Note that for any positive $n, m, l$ and any $z_0^m, z_{-n-l}^{-n}$, we have
$$
p(z_0^m|z_{-n-l}^{-n})=\sum_{z_{-n+1}^{-1}} \frac{\pi \Delta_{z_{-n-l}^{m}} \mathbf{1}}{\pi \Delta_{z_{-n-l}^{-n}} \mathbf{1}} =\frac{\pi \Delta_{z_{-n-l}^{-n}}}{\pi \Delta_{z_{-n-l}^{-n}} \mathbf{1}} (\sum_{z \in \mathcal{Z}} \Delta_z)^{n-1} \Delta_{z_0^m} \mathbf{1}=\frac{\pi \Delta_{z_{-n-l}^{-n}}}{\pi \Delta_{z_{-n-l}^{-n}} \mathbf{1}} \Delta^{n-1} \Delta_{z_0^m} \mathbf{1}.
$$
Let $\lambda_2$ denote the second largest (in modulus) eigenvalue of $\Delta$. By the Perron-Frobenius theory (see, e.g.,~\cite{se80}), $|\lambda_2| < 1$; furthermore, for any $\lambda$ with $|\lambda_2| < \lambda < 1$, there exists $C_1 > 0$ such that for any probability vector $x$, we have
$$
|x \Delta^n-\pi| \leq C_1 \lambda^n.
$$
It then follows that
$$
p(z_0^m|z_{-n-l}^{-n})=\pi \Delta_{z_0^m} \mathbf{1}+O(\lambda^n) \Delta_{z_0^m} \mathbf{1} = p(z_0^m)+O(\lambda^n) p(z_0^m).
$$
Noting that the constant in $O(\lambda^n)$ is independent of $n, m, l$ and $z_0^m, z_{-n-l}^{-n}$, we then conclude that for any $U \in \mathcal{B}(Z_{-\infty}^{-n}), V \in \mathcal{B}(Z_{0}^{\infty})$,
$$
P(V|U)=P(V)+O(\lambda^n)P(V),
$$
which immediately implies the lemma.
\end{proof}

In the following, we shall establish the main results by invoking the limit theorems in Section~\ref{Limit-Theorems}. Before doing so, we set
\begin{equation} \label{set-1}
f(Z_i|Z_1^{i-1})=D^l_{\theta} \log p(Z_i|Z_1^{i-1}),  \qquad X_i = f(Z_i|Z_1^{i-1})-E_{\theta_0}[f(Z_i|Z_1^{i-1})],
\end{equation}
and
\begin{equation} \label{set-2}
S_n=\sum_{i=1}^n X_i, \qquad \sigma_n^2=Var(S_n), \qquad \sigma=\lim_{n \to \infty} \sigma_n^2/n.
\end{equation}
Then, by Corollary~\ref{Real-ExpoForgetting} and Lemma~\ref{Mixing}, Conditions (a), (b) and (c) are satisfied.

\subsection{Proof of Theorem~\ref{HMM-LLL-Theorem}}

Note that for any $i \geq j$, applying Corollary~\ref{Real-ExpoForgetting}, we have
$$
E_{\theta_0}[D^l_{\theta} \log p^{\theta}(Z_i|Z^{i-1}_{j})]-E_{\theta_0}[D^l_{\theta} \log p^{\theta}(Z_i|Z^{i-1}_{j-1})] = O(\rho^{i-j}),
$$
which implies that as $j \to -\infty$, $E_{\theta_0}[D^l_{\theta} \log p^{\theta}(Z_i|Z^{i-1}_{j})]$ converges to a limit, say $E_{\theta_0}[D^l_{\theta} \log p^{\theta}(Z_i|Z^{i-1}_{-\infty})]$, such that
$$
E_{\theta_0}[D^l_{\theta} \log p^{\theta}(Z_i|Z^{i-1}_{j})]-E_{\theta_0}[D^l_{\theta} \log p^{\theta}(Z_i|Z^{i-1}_{-\infty})] = O(\rho^{i-j}).
$$
It then follows that
\begin{align*}
\frac{E_{\theta_0}[D^l_{\theta} \log p^{\theta}(Z_1^n)]}{n} & =\frac{\sum_{i=1}^n E_{\theta_0}[D^l_{\theta} \log p^{\theta}(Z_i|Z^{i-1}_1)]}{n} \\
&=\frac{\sum_{i=1}^n (E_{\theta_0}[D^l_{\theta} \log p^{\theta}(Z_i|Z^{i-1}_{-\infty})]+O(\rho^{i}))}{n}
\end{align*}
which converges to $E_{\theta_0}[D^l_{\theta} \log p^{\theta}(Z_0|Z^{-1}_{-\infty})]$ as $n$ tends to infinity. This implies the well-definedness of $L^{(l)}(\theta)$.

Now, with (\ref{set-1}) and (\ref{set-2}), invoking Theorem~\ref{LLL-Theorem}, we have
$$
\frac{D^l_{\theta} \log p^{\theta}(Z_1^n)}{n}-\frac{E_{\theta_0}[D^l_{\theta} \log p^{\theta}(Z_1^n)]}{n} \to 0 \mbox{ as } n \to \infty,
$$
which, by the definition of $L^{(l)}(\theta)$, implies the theorem.

\subsection{Proof of Theorem~\ref{HMM-CLT-Theorem}}

We will need the following lemma, whose proof follows from Corollary~\ref{Real-ExpoForgetting} and Lemma~\ref{Mixing} and a completely parallel argument as in the proof of Lemma~\ref{VarianceLemma}, and thus omitted.
\begin{lem} \label{HMM-VarianceLemma}
Assume Conditions (I) and (II) and consider a compact subset $\Omega_0 \subset \Omega$ and any $l \geq 0$. For any $0 < \eps_0 < 1$, there exists $C > 0$ such that for any $m, n$ and any $\theta \in \Omega_0$,
$$
\left|\frac{(\sigma_n^{(l)}(\theta))^2}{n}-(\sigma^{(l)}(\theta))^2 \right| \leq C n^{-\eps_0}.
$$
\end{lem}
\noindent Lemma~\ref{HMM-VarianceLemma} immediately implies that
\begin{equation} \label{bound-4}
|\sigma_n^{(l)}(\theta)|=\Theta(\sqrt{n}),
\end{equation}
and furthermore, for any small $\eps_0 > 0$, any $m, n$ and any $\theta \in \Omega_0$,
\begin{equation} \label{bound-1}
|\sigma_n^{(l)}(\theta)- \sqrt{n} \sigma^{(l)}(\theta)| = O(n^{-1/2+\eps_0}).
\end{equation}
Notice that by Corollary~\ref{Real-ExpoForgetting},
\begin{equation} \label{bound-2}
E_{\theta_0}[D_{\theta}^l \log p(Z_1^n)]-n L^{(l)}(\theta) = O(\sum_{i=1}^{\infty} \rho^i)=O(1),
\end{equation}
and by Lemma~\ref{uniformly-bounded},
\begin{equation} \label{bound-3}
|D_{\theta}^l \log p(Z_1^n)-E_{\theta_0}[D_{\theta}^l \log p(Z_1^n)]|=O(n).
\end{equation}
Applying (\ref{bound-1}), (\ref{bound-2}), (\ref{bound-3}) and (\ref{bound-4}), we then have,
for some $\eps_0 > 0$ sufficiently small,
$$
\left|\frac{D_{\theta}^l \log p(Z_1^n)-n L^{(l)}(\theta)}{\sqrt{n} \sigma^{(l)}(\theta)}-\frac{D_{\theta}^l \log p(Z_1^n)-E_{\theta_0}[D_{\theta}^l \log p(Z_1^n)]}{\sigma_n^{(l)}(\theta)} \right|
$$
$$
\leq \left|\frac{D_{\theta}^l \log p(Z_1^n)-n L^{(l)}(\theta)}{\sqrt{n} \sigma^{(l)}(\theta)}-\frac{D_{\theta}^l \log p(Z_1^n)-E_{\theta_0}[D_{\theta}^l \log p(Z_1^n)]}{\sqrt{n} \sigma^{(l)}(\theta)} \right|
$$
$$
+\left|\frac{D_{\theta}^l \log p(Z_1^n)-E_{\theta_0}[D_{\theta}^l \log p(Z_1^n)]}{\sqrt{n} \sigma^{(l)}(\theta)}-\frac{D_{\theta}^l \log p(Z_1^n)-E_{\theta_0}[D_{\theta}^l \log p(Z_1^n)]}{\sigma_n^{(l)}(\theta)} \right|
$$
\begin{equation} \label{bound-5}
\hspace{-1cm} =\left|\frac{E_{\theta_0}[D_{\theta}^l \log p(Z_1^n)]-n L^{(l)}(\theta)}{\sqrt{n} \sigma^{(l)}(\theta)} \right|
+|D_{\theta}^l \log p(Z_1^n)-E_{\theta_0}[D_{\theta}^l \log p(Z_1^n)]| \frac{|\sigma_n^{(l)}(\theta)-\sqrt{n} \sigma^{(l)}(\theta)|}{\sqrt{n} \sigma^{(l)}(\theta) \sigma_n^{(l)}(\theta)} =O(n^{-1/2+\eps_0}).
\end{equation}
Finally, with (\ref{set-1}) and (\ref{set-2}), invoking Theorem~\ref{CLT-Theorem}, we have
$$
P\left(\frac{D_{\theta}^l \log p(Z_1^n)-n L^{(l)}(\theta)}{\sqrt{n} \sigma^{(l)}(\theta)} < x \right)
$$
$$
\hspace{-2cm} =P\left(\frac{D_{\theta}^l \log p(Z_1^n)-E_{\theta_0}[D_{\theta}^l \log p(Z_1^n)(\theta)]}{\sigma_n^{(l)}(\theta)} < x + \frac{D_{\theta}^l \log p(Z_1^n)-E_{\theta_0}[D_{\theta}^l \log p(Z_1^n)(\theta)]}{\sigma_n^{(l)}(\theta)}-\frac{D_{\theta}^l \log p(Z_1^n)-n \sigma_n^{(l)}(\theta)}{\sqrt{n} \sigma^{(l)}(\theta)}  \right)
$$
$$
=G\left(x + \frac{D_{\theta}^l \log p(Z_1^n)-E_{\theta_0}[D_{\theta}^l \log p(Z_1^n)(\theta)]}{\sigma_n^{(l)}(\theta)}-\frac{D_{\theta}^l \log p(Z_1^n)-n \sigma_n^{(l)}(\theta)}{\sqrt{n} \sigma^{(l)}(\theta)} \right)+O(n^{-1/4+\eps_0})
$$
It then follows from (\ref{bound-5}) that for any small $\eps_0 >0$
$$
P\left(\frac{D_{\theta}^l \log p(Z_1^n)-n L^{(l)}(\theta)}{\sqrt{n} \sigma^{(l)}(\theta)} < x \right)=G(x) + O(n^{-1/2+\eps_0}) +O(n^{-1/4+\eps_0})=G(x) + O(n^{-1/4+\eps_0}).
$$
We then have established the theorem.

\subsection{Proof of Theorem~\ref{HMM-ASIP-Theorem}}

With (\ref{set-1}) and (\ref{set-2}), invoking Theorem~\ref{ASIP-Theorem}, we can redefine the process $\{S(t), t \geq 0\}$ on a richer probability space together  with the standard Brownian motion $\{B(t), t \geq 0\}$ such that for any $\eps > 0$,
$$
\sum_{n \leq t} D^l_{\theta} \log p^{\theta}(Z_1^n)-\sum_{n \leq t} E_{\theta_0}[D^l_{\theta} \log p^{\theta}(Z_1^n)]-B((\sigma^{(l)}(\theta))^2 t)=O(t^{1/3+\eps}) \mbox{ a.s. as $t \to \infty$}.
$$
The theorem then follows from (\ref{bound-2}).

\subsection{Proof of Theorem~\ref{HMM-LIL-Theorem}}

With (\ref{set-1}) and (\ref{set-2}), invoking Theorem~\ref{LIL-Theorem}, we have
$$
\limsup_{n \to \infty} \frac{D^l_{\theta} \log p^{\theta}(Z_1^n)-E_{\theta_0}[D^l_{\theta} \log p^{\theta}(Z_1^n)]}{(2n (\sigma^{(l)}(\theta))^2 \log \log n (\sigma^{(l)}(\theta))^2)^{1/2}}=1 \qquad {a.s.}
$$
The theorem then follows from (\ref{bound-2}).

\subsection{Proof of Theorem~\ref{HMM-Chernoff-Bound-Theorem}}

With (\ref{set-1}) and (\ref{set-2}), invoking Theorem~\ref{Chernoff-Bound-Theorem}, we deduce
that for any $x, \eps > 0$, there exist $C > 0$, $0 < \gamma < 1$ such that
$$
P\left(\frac{D^l_{\theta} \log p^{\theta}(Z_1^n)-E_{\theta_0}[D^l_{\theta} \log p^{\theta}(Z_1^n)]}{n} \geq x \right) = O(\gamma^{n^{1-\eps}}).
$$
The theorem then follows from (\ref{bound-2}).

\subsection{Proof of Theorem~\ref{HMM-MLE}}

Again, in this proof, we treat $\theta$ as a one dimensional variable; without loss of generality, we further assume that $L^{(2)}(\theta) > 0$ for all $\theta \in \Omega_0$.

By the mean value theorem, for any $\theta_n$, there exists a $\bar{\theta}_n$, a convex combination of $\theta_0$ and $\theta_n$, such that
$$
D_{\theta} \log p^{\theta_n}(Z_1^n)=D_{\theta} \log p^{\theta_0}(Z_1^n)+D_{\theta}^2 \log p^{\bar{\theta}_n}(Z_1^n)(\theta_n-\theta_0).
$$
And, by the definition of $\theta_n$,
$$
D_{\theta} \log p^{\theta_n}(Z_1^n)=0.
$$
It then follows that for any $x > 0$
$$
P(|\theta_n-\theta_0| \geq x)=P\left(\left|\frac{D_{\theta} \log p^{\theta_0}(Z_1^n)/\sqrt{n}}{D_{\theta}^2 \log p^{\bar{\theta}_n}(Z_1^n)/n}\right| \geq \sqrt{n} x \right).
$$
It follows from negativity of the relative entropy~\cite{co06} that for all $\theta \in \Omega$ and for all $n$,
$$
E_{\theta_0}[\log p^{\theta}(Z_1^n)] \leq E_{\theta_0}[\log p^{\theta_0}(Z_1^n)],
$$
which implies that
$$
E_{\theta_0}[D_{\theta} \log p^{\theta_0}(Z_1^n)]=0, \mbox{ and thus } L^{(1)}(\theta_0)=0.
$$
Then, by Theorem~\ref{HMM-Chernoff-Bound-Theorem}, for any $x_1, \eps_1 > 0$, there exist $C_1 > 0$, $0 < \gamma_1 < 1$ such that for any $n$ and any $\theta \in \Omega_0$,
$$
P\left( F(x_1) \right) \leq C_1 \gamma_1^{n^{1-\eps_1}},
$$
where $F(x_1)$ denotes the event that
$$
\left|\frac{D^2_{\theta} \log p^{\theta}(Z_1^n)}{n}-L^{(2)}(\theta) \right| \geq x_1.
$$
By Theorem~\ref{HMM-CLT-Theorem}, for any $x_1, \eps_2 > 0$, there exists $C_2 > 0$ such that
$$
P(|\theta_n-\theta_0| \geq x, F^c(x_1)) \leq P(|\mathcal{N}| \geq (L^{(2)}(\bar{\theta}_n)-x_1) \sqrt{n} x)+ C_2 n^{-1/4+\eps_2},
$$
where $\mathcal{N}$ denotes the standard normal random variable. It then follows that
\begin{align*}
P(|\theta_n-\theta_0| \geq x) & = P(|\theta_n-\theta_0| \geq x, F^c(x_1))+ P(|\theta_n-\theta_0| \geq x, F(x_1))\\
& \leq P(|\mathcal{N}| \geq (L^{(2)}(\bar{\theta}_n)-x_1) \sqrt{n} x)+ C_2 n^{-1/4+\eps_2}+ P(F(x_1))\\
& \leq e^{-n (L^{(2)}(\bar{\theta}_n)-x_1)^2}+C_2 n^{-1/4+\eps_2}+C_1 \gamma_1^{n^{1-\eps_1}},
\end{align*}
where we have used the fact that for any $y > 0$
$$
P(|\mathcal{N}| \geq y) \leq e^{-y^2}.
$$
The theorem then immediately follows if we choose $x_1 > 0$ sufficiently small such that for all $\theta \in \Omega_0$,
$$
L^{(2)}(\theta)-x_1 > 0.
$$

%\bigskip
%
%\textbf{Acknowledgement.} We would like to thank Brian Marcus, Venkat Anantharam, Yury Polyanskiy and Dongning Guo for very helpful discussions.

\end{document}